\documentclass{jpp}

\usepackage{graphicx}
\usepackage{epstopdf}
\usepackage{xcolor}
\usepackage{amsmath}
\usepackage[export]{adjustbox}

\usepackage{xspace}
\newcommand\ie{\textit{i.e.}\xspace}
\usepackage{comment}

\newcommand{\vect}[1]{\boldsymbol{#1}}

\def \Ba {B_{\mathrm{a}}}

\providecommand\bnabla{\boldsymbol{\nabla}}
\providecommand\bkappa{\boldsymbol{\kappa}}

\def \Bcs {\bar{B}}

\usepackage{unicode}
\newcommand{\csso}{\fontencoding{LECO}\selectfont\char215}
\newcommand{\iotaslash}{\hspace*{0.1em}\iota\hspace*{-0.45em}\text{\csso}}

\providecommand\bnabla{\boldsymbol{\nabla}}
\providecommand\bkappa{\boldsymbol{\kappa}}

\title{Direct construction of optimized stellarator shapes.  III.  Omnigenity near the magnetic axis.}
\author{Gabriel G. Plunk\aff{1}\corresp{\email{gplunk@ipp.mpg.de}}, Matt Landreman\aff{2}, Per Helander\aff{1}
  }
\affiliation{\aff{1}Max-Planck-Institut f\"ur Plasmaphysik, EURATOM Association, 17491 Greifswald, Germany
\aff{2}Institute for Research in Electronics and Applied Physics, University of Maryland, College Park MD 20742, USA
}

\begin{document}
\maketitle

\begin{abstract}
The condition of omnigenity is investigated, and applied to the near-axis expansion of \citet{garren-boozer-1}.  Due in part to the particular analyticity requirements of the near-axis expansion, we find that, excluding quasi-symmetric solutions, only one type of omnigenity, namely quasi-isodynamicity, can be satisfied at first order in the distance from the magnetic axis.  Our construction provides a parameterization of the space of such solutions, and the cylindrical reformulation and numerical method of \citet{landreman-sengupta, landreman-sengupta-plunk}, enables their efficient numerical construction.
\end{abstract}

\section{Introduction}

The ``near axis expansion'', an expansion in the distance from the magnetic axis, has been investigated as a method to characterize the space of magnetic equilibria, and determine how well the condition of quasi-symmetry can be maintained across the plasma volume \citep{garren-boozer-1, garren-boozer-2}.  More recently, it is being reconsidered as a method to rapidly generate such equilibria, without resorting to costly numerical optimization.

The present work, the third paper in a series \citep{landreman-sengupta, landreman-sengupta-plunk}, is concerned with extending these ideas to the broader class of omnigenous fields, which confine collisionless particle orbits, but do not necessarily satisfy quasi-symmetry.  By relaxing this requirement, a large amount of freedom is gained, and the solution space can be considered as an exhaustive catalog of ``good'' equilibria (\ie all those that confine trapped particle orbits).  We note that some analytical work of this type provided a basis for the development of stellarator configurations \citep{gori-lotz-nuehrenberg, nuerenberg-ppcf-2010}, but was limited to treatment of deeply trapped particles in a particular (quasi-isodynamic) class of equilibria.

Two related properties, characterizing optimal stellarator equilibria, are quasi-symmetry and omnigenity.  The relationship between them is somewhat paradoxical, as found by \citet{cary-shasharina}:  On the one hand, quasi-symmetry is formally a subclass of omnigenity -- all quasi-symmetric fields are also omnigenous.  However, those omnigenous fields which break quasi-symmetry are, under reasonable conditions, also non-analytic, so that under the restriction of analyticity, it appears, disappointingly, that these classes are the same.  However, replacing a non-analytic solution with a sufficiently smooth approximation, one can indeed approach omnigenity while sharply violating quasi-symmetry, explaining why it is indeed fruitful to simply search for equilibria that confine particles, without necessarily imposing the satisfaction of quasi-symmetry.

These facts set the stage for our near-axis expansion, where analyticity is indeed assumed in the two dimensions perpendicular to the magnetic axis.  It should be no surprise, therefore, that we find two variants of omnigenity (those with toroidally and helically closed contours of magnetic field strength) cannot be achieved without breaking quasi-symmetry -- one, however, cannot rule out their existence entirely, as they might be found via a less restrictive construction.

We do, however, find that omnigenity can be approximately satisfied (at first order in the distance from the magnetic axis) for the class of equilibria with polloidally closed contours of magnetic field strength, so-called quasi-isodynamic equilibria.  Constructing such equilibria presents some technical difficulties (as compared with quasi-symmetric equilibria), but the potential benefit in terms of freedom is quite substantial -- three additional free functions of the toroidal angle, as compared with quasi-symmetric construction!

The near-axis construction of optimal stellarator equilibria can thus be summarized as follows.  Quasi-axisymmetric and quasi-helically symmetric equilibria can be directly constructed, but there are no other consistent omnigenous equilibria with toroidally or helically closed contours of field strength.  On the other hand, one cannot construct quasi-polloidally symmetric equilibria near the magnetic axis, but we find that one can indeed find approximately quasi-isodynamic ones, \ie quasi-symmetry {\it must} be broken to achieve this kind of omnigenity.  Therefore, it seems that quasi-helical symmetry, quasi-axial symmetry, and quasi-isodynamicity are the three natural ways to achieve good stellarator equilibria with a near-axis approximation.  These are the only types of equilibria with collisionless orbit confinement and therefore small neoclassical transport. It should be noted, however, that quasi-isodynamic equilibria do not share other neoclassical properties of quasisymmetric ones. The bootstrap current vanishes identically \citep{Helander_2009,landreman-catto} and the transport is not intrinsically ambipolar, which severely restricts plasma rotation \citep{helander-pop, helander-simakov}.

The paper is organized as follows.  In Sec.~\ref{sec:omnigenity-def} we review the definition and geometric interpretation of omnigenity, and introduce a coordinate mapping, based on that of \citet{cary-shasharina}.  We use the mapping as a tool to construct omnigenous magnetic field strength functions.  In particular, in Sec.~\ref{sec:omnigenity-near-qs}, the mapping is expanded for nearly quasisymmetric fields, which, due to its invertibility in this case, means that such fields can be parameterized by a free function representing the perturbation.  Although this expansion may have more general application, it is particularly useful for generating equilibria using the near axis expansion, since such equilibria must necessarily be nearly quasi-symmetric.  This is discussed in Sec.~\ref{sec:B-compare}, where the near-axis and omnigenous forms of the magnetic field strength are checked for consistency.  In this section two classes of omnigenity are shown to yield inconsistencies, and are therefore eliminated from consideration, leaving only the third, quasi-isodynamicity.  After demonstrating consistency in the forms of magnetic field strength, the remaining conditions needed to construct an equilibrium solution are then examined in Sec.~\ref{sec:QI-theoretical-construction}.  In Sec.~\ref{sec:numerical-solutions} it is demonstrated numerically that the equilibrium equation can indeed be solved, yielding solutions that satisfy omnigenity to the degree predicted by the theory.  The results are summarized and discussed in Sec.~\ref{sec:conclusion}.

\section{Omnigenity}\label{sec:omnigenity-def}

The most fundamental optimization for a stellarator is the shaping of the magnetic field to confine collisionless trapped particle orbits.  This is achieved when there is no "radial" excursion of such particles during a single orbit, as expressed in terms of the change in the coordinate labelling the magnetic flux surfaces, here $\psi$ ($2\pi \psi$ is the toroidal magnetic flux).  Denoting the magnetic drift velocity of a particle as ${\bf v}_d$ and using $d\psi/dt = {\bf v}_d \cdot \bnabla \psi$, the condition is written as $\Delta \psi = 0$, where, for a given  magnetic well, is found to be

\begin{align}
    \Delta \psi &= \oint dt\; {\bf v}_d \cdot \bnabla \psi,\\
    &= 2\int_{l_1}^{l_2} \frac{dl}{v_\parallel} {\bf v}_d \cdot \bnabla \psi,\\
    &= \sum_\gamma 2\gamma \int_{B_\mathrm{min}}^{1/\lambda} \frac{dB}{v_\parallel \partial B/\partial l} {\bf v}_d \cdot \bnabla \psi,\\
    &= \sum_\gamma \gamma \int_{B_\mathrm{min}}^{1/\lambda} dB \; h(B, \lambda)\; Y(\psi, \alpha, B, \gamma).\label{Delta-psi-4}
\end{align}
In the first line, the quantity is represented as a time average.  In the second line, it is rewritten as an integral over the arc length $l$ between two bounce points $l_1$ and $l_2$.  In the third line, the arc length is replaced by the magnetic field strength as a variable of integration, and the sign $\gamma \equiv \mathrm{sign}(\partial B/\partial l)$ is introduced which identifies the two sides of the magnetic well, and the upper bound of integration is written explicitly in terms of the pitch angle $\lambda = v_\perp^2/(v^2 B)$.  To obtain the final line, we use the expression for the drift velocity of a particle of mass $m$ and charge $q$, ${\bf v}_d = \frac{m\hat{\boldsymbol b}}{q B}\times(v_\parallel^2 \bkappa + \frac{v_\perp^2}{2}\bnabla \ln B)$, where $\bkappa = \hat{\boldsymbol b}\cdot\bnabla\hat{\boldsymbol b}$ is magnetic curvature, $\hat{\boldsymbol b} = {\boldsymbol B}/B$, and note that $\bnabla\psi \cdot ({\bf B}\times \bkappa) = \bnabla\psi \cdot ({\bf B}\times \bnabla\ln B)$.  Thus, the integrand may be separated into a purely geometric factor $Y$ (depending generally on the flux surface label $\psi$, field line label $\alpha$, and the magnetic field strength),

\begin{equation}
    Y = \frac{\bnabla \psi \times {\bf B} \cdot \bnabla B}{{\bf B}\cdot\bnabla B},\label{Y-def}
\end{equation}
 and an additional factor $h =\frac{2mv}{q B}\frac{1-\lambda B/2}{\sqrt{1-\lambda B}}$ depending on the magnetic field strength and the pitch angle.  The expression for $\Delta \psi$ given by \ref{Delta-psi-4} can be viewed as an integral transformation from $B$ space to $\lambda$ space (the velocity $v$ appears only as a multiplicative factor), which can be inverted, implying that the condition $\Delta \psi = 0$ is satisfied if and only if

\begin{equation}
    \sum_\gamma \gamma Y = 0.\label{omnigenity-condition}
\end{equation}
For a quasisymmetric field, $Y$ is constant on flux surfaces \citep{helander-simakov} and this condition is trivially satisfied.  Eqn.~\ref{omnigenity-condition} is a purely geometric condition, in the sense that it is a condition on the spatial dependence of the magnetic field strength.  For the purposes of this paper, we will consider the consequences of this condition on the magnetic field in Boozer poloidal and toroidal coordinates, $\theta$ and $\varphi$ respectively.  Using Eqn.~\ref{Y-def}, we write the magnetic field in the numerator in the covariant form ${\bf B}_{\mathrm{cov}} = G\bnabla\varphi + I \bnabla \theta + K \bnabla \psi$, and the field in contravariant form in the denominator, ${\bf B}_{\mathrm{con}} = \bnabla\psi\times\bnabla(\theta - \iotaslash\varphi)$, obtaining

\begin{equation}
    Y = \frac{I(\psi) \frac{\partial B}{\partial \varphi} - G(\psi) \frac{\partial B}{\partial \theta}}{\frac{\partial B}{\partial \varphi} + \iotaslash(\psi) \frac{\partial B}{\partial \theta}},
\end{equation}
where the toroidal and poloidal currents $I(\psi)$ and $G(\psi)$ have been introduced.  This expression can be interpreted as the ratio of two (linearly independent) components of the gradient of B in the $\theta$-$\varphi$ plane.  Eqn.~\ref{omnigenity-condition} states that this ratio must be equal at the two (bounce) points of equal magnetic field strength, \ie $\gamma = \pm 1$.  This fact has the simple geometric interpretation that the gradients of $B$ must be anti-parallel, \ie the contours of constant magnetic field strength must be parallel at those two points.  Equivalently, the angular separation between points of equal magnetic field strength (within a given magnetic well) must be the same for all field lines.  A number of corollaries follow intuitively, for instance that the contours of magnetic field strength must all close in the same sense (either poloidally, toroidally, or helically), and the contour of the global maximum must be straight (for irrational rotational transform).

\section{Cary-Shasharina (CS) mapping}

It was shown by \citet{cary-shasharina} that the condition of omnigenity can be satisfied by functions $B(\xi, \zeta)$ that satisfy a set of geometric conditions.  Here, $\xi$ and $\zeta$ are angular coordinates (defined in terms of Boozer angles) defined so that contours of constant magnetic field strength close in the $\xi$ direction.  We will use the angle $\xi$ to denote either either the poloidal, or toroidal coordinate, $\theta$ or $\varphi$, depending on the class of symmetry.  The idea with the CS mapping is to introduce a new coordinate $\eta$, to replace the $\zeta$ coordinate, so that lines of constant magnetic field strength will appear straight in the $\eta$-$\xi$ plane.

The expression for magnetic field strength given by \citet{cary-shasharina} is

\begin{equation}
B = B_0(1 + \epsilon_r \cos(\eta)),\label{CS-B-fcn}
\end{equation}
where $|\epsilon_r| < 1$ ensures that $B$ has no zeros; other expressions would also work, but this one has the benefit that it is analytic in $\eta$.  The coordinate mapping for $\eta$ is defined via

\begin{equation}
\zeta = \eta + F(\xi, \eta).\label{CS-map-exact}
\end{equation}
When expressed in this manner, an explicit form of $F$ can be given, from which the condition of omnigenity is enforced

\begin{equation}\label{F-CS}
F(\xi, \eta) = \begin{cases}
f(\xi, \eta),\quad \text{for } 0 \leq \eta \leq \pi, \\
2\pi - 2\eta + f(\xi - \iotaslash \Delta \zeta , 2\pi - \eta) + \Delta \zeta, \quad \text{for } \pi \leq \eta \leq 2\pi,
\end{cases}
\end{equation}
where we note that $\iotaslash$ is the rotational transform in the $\xi$-$\zeta$ plane (number of $\zeta$ transits per $\xi$ transit).  The contour of maximum field strength is straight, {\em i.e.} $\eta = 0$ at $\zeta = 0$, independent of $\xi$, so we have the additional constraint

\begin{equation}
f(\xi, 0) = 0.
\end{equation}
The function $\Delta \zeta(\eta)$ specifies the angular separation, as measured along a magnetic field line, between a contour of constant $\eta$, say $\eta = \eta_0$, and the corresponding contour where the magnetic field has the same strength, \ie $\eta_1 = 2\pi - \eta_0$; this function is therefore defined for $\pi \leq \eta \leq 2\pi$, and has the properties $\Delta \zeta(\pi) = 0$ and $\Delta\zeta(2\pi) = 2\pi$.  The condition of omnigenity is that this angular separation only depends on the magnetic field strength, in this case labeled by $\eta$.  

The above construction only considers the relatively simple case of a single magnetic well.  \cite{parra-et-al} demonstrated that more complicated fields are possible, by providing an explicit construction for a two-well case.  It is not surprising that more general omnigenous magnetic fields, with arbitrarily complex arrangement of magnetic wells are also allowed, and a general construction is given in the following section.

\section{General omnigenous fields}\label{sec:gen-omnigenity}

One can think of the CS mapping as representing the general deformation of a one-dimensional prototype magnetic field strength function, \ie Eqn.~\ref{CS-B-fcn}, such that quasi-symmetry is broken while preserving omnigenity.  Let us introduce a general function $\Bcs(\eta)$, and propose a procedure to continuously deform it.  This function is assumed to satisfy 

\begin{equation}
\Bcs(\eta) > 0,\label{B-eta}
\end{equation}
be $2\pi$ periodic, and reach its global maximum at $\eta = 0$.  We would also like this function to possess some degree of smoothness (so that smoothness imparted in the mapping for $\eta$ may be inherited by the magnetic field itself) but it need not be analytic; for instance piece-wise polynomials may be convenient.

\begin{figure}
\centering
\includegraphics[width=\textwidth]{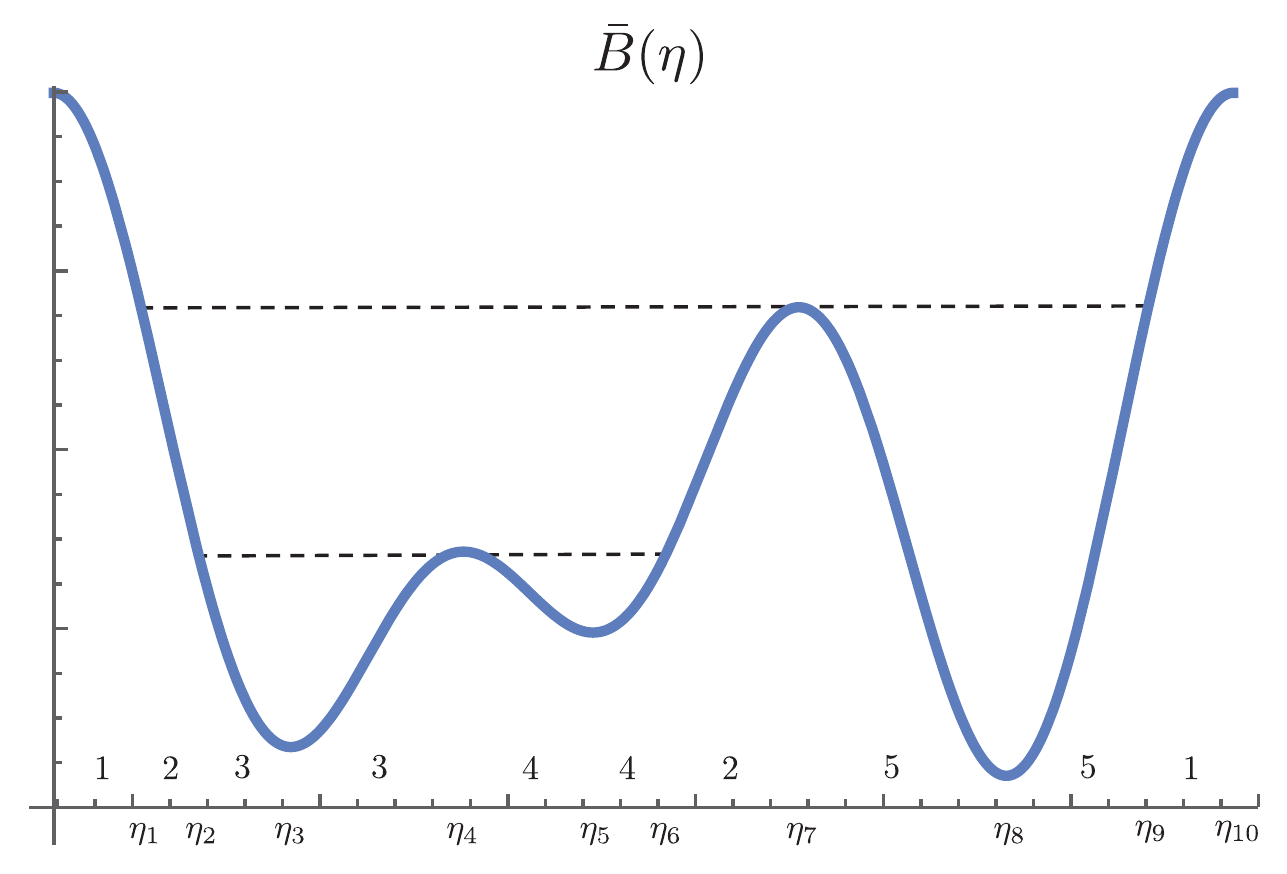}
\caption{Example magnetic field with several magnetic wells.  The numbered intervals can be identified as the trapping domains, \ie from left to right we have $D_{1L}$, $D_{2L}$, $D_{3L}$, $D_{3R}$, $D_{4L}$, $D_{4R}$, $D_{2R}$, $D_{5L}$, $D_{5R}$, $D_{1R}$.}
\label{B-example-fig}
\end{figure}

The function $\Bcs(\eta)$, an example of which is given in Fig.~\ref{B-example-fig}, implies a set of boundary points in $\eta$, which define what we will call trapping domains.  Let us itemize these points:  First there are the locations of local minima and maxima, starting with the global maximum at $\eta = 0$.  Then, each local maximum that is less than the global maximum implies a number of additional boundary points associated with values of $\eta$ where the magnetic field reaches the level of that local maximum.  We will denote these boundary points, in the order they appear along the field line,  as $\eta_0, \eta_1, \eta_2 \dots$, and we will number the trapping domains, defined by neighboring points, in the order of increasing $\eta$, but assign the same index to the two domains that have matching bounce points; see Fig.\ref{B-example-fig}.  Thus, the left-hand domain, where $\Bcs^\prime < 0$ will be numbered $i$, and denoted $D_{iL}$ while the complimenting domain, with $\Bcs^\prime > 0$ will have the same index and be denoted $D_{iR}$.  We denote the unions of left and right trapping domains as $D_L$ and $D_R$.

Next a function $\eta_{b}(\eta, i)$ must be constructed, which returns the left bounce point for a given point in a right-hand domain.  That is, for any point $\eta$ contained in a right-hand domain $D_{iR}$, the function $\eta_b(\eta,i )$ returns the corresponding bounce point in the left-hand part of domain $D_{iL}$.  (The quantity corresponding to $\eta_b$ in the original CS mapping is $2\pi - \eta$.)  Note the explicit dependence on the index $i$, since we must distinguish between domains at locations that exactly coincide with boundary points.  For example, for the case depicted in Fig.~\ref{B-example-fig}, $\eta_{b}(\eta_6, 4) = \eta_4$, while $\eta_{b}(\eta_6, 2) = \eta_2$.  Note that it may be advantageous to use a simple method for constructing $\Bcs(\eta)$, so that the function $\eta_b$ might be defined analytically.

As before, the mapping is given by Eqn.~\ref{CS-map-exact}, but now the function $F$ must be defined over the various trapping domains:

\begin{equation}\label{F-CS-general-1}
F(\xi, \eta) = \begin{cases}
f(\xi, \eta),\quad \text{for } \eta \in D_{iL}, \\
\Delta\zeta + \eta_b - \eta + f(\xi - \iotaslash \Delta \zeta, \eta_{b}), \quad \text{for } \eta \in D_{iR},
\end{cases}
\end{equation}
where $\Delta\zeta(\eta, i)$ prescribes the angular difference between field lines in magnetic coordinate $\zeta$.  Note that this can be confirmed by by evaluating Eqn.~\ref{CS-map-exact} at two bounce points in a well, and taking the difference. That is, take ($\xi_1$, $\zeta_1$, $\eta_1$) in $D_{iL}$ and ($\xi_2$, $\zeta_2$, $\eta_2$) in $D_{iR}$.  Since these are bounce points, we have the identities $\eta_b(\eta_2, i) = \eta_1$, and $\xi_2 - \iotaslash \Delta\zeta(\eta_2, i) = \xi_1$.  The difference of the two equations thus gives $\zeta_2 - \zeta_1 = \Delta\zeta(\eta_2, i)$.

While Eqn.~\ref{F-CS-general-1} is a relatively simple and direct generalization of the CS mapping, it is cumbersome to enforce continuity of $\zeta$ at the boundary points.  In part this is due to the fact that the function $\Delta\zeta$ must be constructed in a consistent manner, for instance so that the angular distance spanning a well is consistent with the distance prescribed in the external trapping regions, resulting in a hierarchy of additional constraints.

Therefore, we propose a simplification, without loss of generality, namely to consider the angular distance as being prescribed by $\Bcs(\eta)$ itself.  That is, we take $\Delta\zeta(\eta, i) = \Delta\eta(\eta, i) = \eta - \eta_b(\eta, i)$.  We arrive then at the simple form

\begin{equation}\label{F-CS-general-2}
F(\xi, \eta) = \begin{cases}
f(\xi, \eta),\quad \text{for } \eta \in D_{iL}, \\
f(\xi - \iotaslash \Delta \eta, \eta_{b}), \quad \text{for } \eta \in D_{iR}.
\end{cases}
\end{equation}

Note now that the function $f(\xi, \eta)$ need only be defined over the left-hand domain $D_L$.  We emphasize that there is no loss of generality in this choice of $\Delta\zeta$, as angular separation between bounce points can be controlled during the construction of $\Bcs(\eta)$.  This choice also has the benefit of separating out the freedom associated with the corresponding narrower class of quasi-symmetric magnetic fields, from the freedom associated with breaking quasi-symmetry.

What remains is merely to enforce continuity of $\zeta$ (thus $F$) at boundary points (one may also wish to enforce continuity of some number of derivatives).  Some of these conditions are trivial: by construction, continuity of $F$ is already satisfied at local minima ($\Delta \eta = 0$ at such points).  Also, at boundary points between two left-hand domains, $F$ will be continuous as long as $f$ is continuous.  The other boundary points must satisfy shifted conditions due to the translation by $- \iotaslash \Delta \eta$ in the $\xi$ coordinate, but this can clearly be satisfied since the function $f$ is completely arbitrary, aside from the requirement that $f(\xi, 0) = 0$ and the invertibility requirement $\partial F/\partial \eta > - 1$ (\ie that the Jacobian of the variable transformation is non-zero).

Thus it is demonstrated that arbitrarily complex magnetic well structures are compatible with omnigenous $B$, and the construction of such functions is only constrained by a set of matching conditions, and an invertibility requirement.  In the remainder of the paper, the definition for $F$ given in Eqn.~\ref{F-CS-general-2} will be used.

\subsection{Analyticity}

Analyticity of the magnetic field $B$ in its spatial coordinates, will, under reasonable conditions, imply analyticity of the function $F$ in the coordinate $\eta$ \citep{Krantz-Parks}.  Using their mapping, \citet{cary-shasharina} showed that (assuming irrational rotational transform) analyticity of $F$ in $\eta$ can only be satisfied for quasi-symmetric fields, in particular continuity of derivatives of the magnetic field at $\eta = 0, 2\pi$, where the field maximum is located, require that the corresponding derivatives of $f$ are zero,

\begin{equation}
\left.\frac{d^n f}{d\eta^n}\right |_{\eta = 0} = 0.\label{cs-analytic-1}
\end{equation}
Thus if $f$ is to be analytic, it must, generally speaking, be independent of $\eta$, \ie it must be quasi-symmetric.  The interesting cases, therefore, are non-analytic.  Note that other boundary points, for instance local minima of the magnetic field (say $\eta_{\mathrm{min}} = \eta_j$) are also locations of potential discontinuity in derivatives.  Therefore, it may be useful to establish the analyticity constraints there as well.  For example, by imposing $(\partial F/\partial \eta)_{\eta_{\mathrm{min}} -} = (\partial F/\partial \eta)_{\eta_{\mathrm{min}} +}$, using Eqn.~\ref{F-CS-general-2}, we arrive at the condition

\begin{equation}
\left[\frac{\partial f}{\partial \eta} = \left(1-\frac{d \Delta \eta}{d\eta}\right)\frac{\partial f}{\partial \eta} - \iotaslash \frac{d \Delta \eta}{d\eta}\frac{\partial f}{\partial \xi}\right]_{\eta = \eta_{\mathrm{min}}}.\label{cs-analytic-2}
\end{equation}
Thus it would seem that analyticity at $\eta_{\mathrm{min}}$ demands a particular relationship between $\Delta\eta$ and $f$ at boundary points.  However, given that analyticity must already be abandoned at $\eta = 0$, it is not obvious why one should demand analyticity at other locations such as $\eta_{\mathrm{min}}$.  Nevertheless, conditions such as \ref{cs-analytic-1} and \ref{cs-analytic-2}, can be used to enforce continuity of arbitrary derivatives of the magnetic field strength, if desired.

\subsection{Invertibility}

It would seem that the mapping \ref{F-CS-general-2} allows one to generate a generic omnigneous field strength using the free functions $f$ and $\Delta \zeta$.  However, a complication arises in that its invertibility is not guaranteed.  That is, although $\zeta$ is uniquely defined in the $\xi$-$\eta$ plane, it is not necessarily true that a unique value of $\eta$ can be obtained as a function of $\xi$ and $\zeta$.  Note, for instance, that two contours of two values of $\eta$ could intersect, which cannot be allowed since the magnetic field strength must be single-valued.

The issue is reminiscent of the problem involving the specification of a general coordinate mapping ${\bf x}(\psi, \theta, \varphi)$, which can generate self-intersecting magnetic surfaces.  Practically, this presents a challenge for the parametrization of the space of omnigneous magnetic field strengths, \ie devising a general method to construct omnigenous functions $B(\xi, \zeta)$.  However, since the change of coordinates is fairly simple (it essentially one-dimensional, $\zeta \rightarrow \eta$), the Jacobian is readily computed, and the condition that it is non-zero could be asserted, \ie

\begin{equation}
\frac{d \zeta}{d \eta } = 1 + \frac{d F}{d\eta} > 0,\quad \text{for all }\xi\text{ and }\eta,
\end{equation}
which translates, via Eqn.~\ref{F-CS-general-2}, into a condition on $f$; note requiring the derivatives of $f$ to be small would suffice.  This might be suitable for numerically generating well-behaved mappings.  However, for present purposes, we find a more convenient approach is to expand about known omnigenous fields, the quasi-symmetric ones, relying on the fact that sufficiently small deviations will preserve the topology of the lines of constant $\eta$, and therefore ensure the invertibility of the mapping.

\section{Omnigenity near quasi-symmetry}\label{sec:omnigenity-near-qs}

In this section we derive an approximate form of omnigenity, expanding about a quasisymmetric field.  Besides the near-axis expansion, this form could have other applications, such as numerical optimization or direct solutions applying omnigenity on a single surface \citep{plunk_helander_2018}.  Let us express the mapping as being composed of a quasi-symmetric and non-quasi-symmetric parts,

\begin{eqnarray}
F = F_0(\eta) + \epsilon F_1(\xi, \eta),\\
f = f_0(\eta) + \epsilon f_1(\xi, \eta).
\end{eqnarray}
The mapping is then defined by

\begin{equation}
\zeta = \eta + F_0(\eta) + \epsilon F_1(\xi, \eta).\label{CS-map-pre-expand}
\end{equation}
We can now consider the deviation to be small, $\epsilon \ll 1$, so that the coordinate $\eta$ may be expanded as

\begin{equation}
\eta = \eta_0 + \epsilon \eta_1.
\end{equation}
The mapping at zeroth order is then

\begin{equation}
\zeta = \eta_0 + F_0(\eta_0).\label{CS-map-0}
\end{equation}
where

\begin{equation}\label{F0-CS}
F_0(\eta) = \begin{cases}
f_0(\eta),\quad \text{for } \eta \in D_{iL} \\
f_0(\eta_b(\eta)),\quad \text{for } \eta \in D_{iR}.
\end{cases}
\end{equation}
Henceforth, and without loss of generality, we will assume $f_0 = 0$, so that $F_0 = 0$ and $\eta_0 = \zeta$.  This amounts to absorbing all the freedom of the zeroth order mapping into the definition of the zeroth order function $\Bcs_0(\eta_0) = \Bcs_0(\zeta)$.  Note that $\Bcs_0$ also determines the function $\eta_b$ and $\Delta \eta$.  The first order terms of Eqn.~\ref{CS-map-pre-expand} yield

\begin{equation}
\eta_1 = -F_1(\xi, \zeta),\label{eta-1-eqn}
\end{equation}
where

\begin{equation}\label{F1-CS}
F_1(\xi, \eta) = \begin{cases}
f_1(\xi, \eta),\quad \text{for } \eta \in D_{iL} \\
f_1(\xi - \iotaslash\Delta\eta(\eta), \eta_b), \quad \text{for }  \eta \in D_{iR}.
\end{cases}
\end{equation}
The magnetic field to first order can be expressed as

\begin{equation}
\Bcs \approx \Bcs_0(\zeta) + \epsilon[\eta_1 \Bcs_0^\prime(\zeta) + \Bcs_1(\zeta)],\label{om-B}
\end{equation}
where we have introduced the small parameter $\epsilon$, and introduced $\Bcs_1$, which can account for changes to the functional form of $\Bcs$ (small modifications of well depths, heights, locations, \etc). 

Thus, the problem of specifying an omnigenous field has been reduced to specifying the zeroth order quasi-symmetric magnetic field, and adding an arbitrary, small deviation to the CS mapping via the perturbation $f_1$.  There is a great deal of freedom in specifying $f_1$: it must only satisfy the appropriate continuity conditions, and $f_1|_{\eta = 0} = 0$, and, optionally, continuity in derivatives of a desired order.

\section{Compatibility of omnigenous and near-axis forms of $B$}\label{sec:B-compare}

In the near-axis expansion, the magnetic field is nearly quasi-symmetric in a trivial sense, because the zeroth order contribution (\ie the on-axis magnetic field) can only depend on the Boozer toroidal angle.  The field is therefore simultaneously quasi-axisymmetric, quasi-helically symmetric and quasi-poloidally symmetric on the axis if the on-axis field is constant; if the field strength varies with the Boozer toroidal angle, it cannot be quasi-axisymmetric or quasi-helically symmetric, but it is still automatically quasi-poloidally symmetric.  At a small distance from the axis, we are concerned with whether omnigenity can still be satisfied, when quasi-symmetry is broken.  The first step is to check for consistency between two forms of the magnetic field strength, namely the omnigenous form and the near-axis form.

In section \ref{const-sec} we consider the case of constant on-axis magnetic field, and conclude that the two forms are inconsistent, implying that omnigenous non-quasi-symmetric equilibria of this type do not exist -- at least not any that are consistent with the axis expansion to first order.  The question of consistency for the case of varying on-axis field strength is then treated in section \ref{sec:quasi-isodynamic}, where more optimistic conclusions are drawn.

Following \citet{garren-boozer-1}, we write the general form of the magnetic field to first order in the distance from the axis

\begin{equation}
B(\epsilon, \theta, \varphi) \approx \Ba(\varphi) \left(1 + \epsilon d(\varphi) \cos[\theta - \alpha(\varphi) ] \right),\label{GB-B}
\end{equation}
where  $\epsilon \ll 1$ can be taken as a measure of the distance from the axis.  Please note that $\alpha(\varphi)$ is not to be confused with the field line label in the Clebsch representation of the magnetic field.  We will compare this form to the general form of a weakly non-quasi-symmetric omnigenous magnetic field, given by Eqn.~\ref{om-B}, for the three symmetry classes.

\subsection{Impossibility of omnigenity with constant on-axis magnetic field}\label{const-sec}

Let us expand the magnetic field $\Bcs(\eta)$, introduced in Eqn.~\ref{B-eta}, about its constant on-axis value:

\begin{equation}
\Bcs \approx \Bcs_0 + \epsilon \Bcs_1(\eta)\label{CS-B-const}
\end{equation}
Note that we do not employ the expansion of section \ref{sec:omnigenity-near-qs} here, which is inapplicable as the first order field $\Bcs_1$ determines the magnetic well structure non-perturbatively when the zeroth order field $\Bcs_0$ is constant.

Within Sec.~\ref{const-sec}, we will allow this to describe a possibly non-omigenous field, that nevertheless does satisfy a more relaxed condition called ``pseudo-symmetry''.  Pseudo-symmetry means that the contours of magnetic field all have a common topology -- \ie they all close poloidally, all close toroidally, or all close helically -- which is a necessary but not sufficient requirement for omnigenity \citep{subbotin, Helander_2011}.  To assert pseudo-symmetry we assume that Eqn.~\ref{CS-map-exact} still applies, but with $F$ not necessarily satisfying Eqn.~\ref{F-CS-general-2}.  Equating the two forms of the magnetic field, Eqns.~\ref{GB-B}-\ref{CS-B-const}, we obtain at dominant order $\Bcs_0 = \Ba$, \ie a constant, and at first order

\begin{equation}
\frac{\Bcs_1(\eta)}{\Ba} = d(\varphi) \cos(\theta - \alpha(\varphi)).\label{CS-B1-const}
\end{equation}

\subsubsection{Toroidally or helically closed lines of magnetic field strength}
Let us consider the cases of toroidally or helically closed contours of constant magnetic field, taking $\zeta = \theta - N \varphi$ and $\xi = \varphi$, where $N = 0$ for the toroidal case.  Let us write $\alpha = \widetilde{\alpha} + N\varphi$, so that Eqn~\ref{CS-B1-const} becomes

\begin{equation}
\Bcs_1(\eta) = d(\varphi) \Ba \cos(\zeta - \widetilde{\alpha}(\varphi)).\label{CS-B1-const-2}
\end{equation}

First, we note that pseudo-symmetric fields are consistent with Eqn.~\ref{CS-B1-const-2}, for instance $\Bcs_1 \propto \cos(\zeta + c \sin(\varphi))$, with $|c| < 1$.  This example is not, however, omnigenous since the contour of maximum field strength (\ie $\eta = 0$) is not straight.  In fact, omnigenous examples are not possible, which can be proved as follows.  First we can determine $d(\varphi)$ by evaluating Eqn.~\ref{CS-B1-const-2} at the global maximum, $\eta = \zeta = 0$, yielding

\begin{equation}
d(\varphi) = \frac{\Bcs_1(0)}{\Ba \cos(\widetilde{\alpha})}
\end{equation}
Eqn.~\ref{CS-B1-const-2} then becomes

\begin{equation}
\Bcs_1(\eta) = \Bcs_1(0) \frac{\cos(\zeta - \widetilde{\alpha})}{\cos(\widetilde{\alpha})}.
\end{equation}
We can show that the only omnigenous fields consistent with this form are quasi-symmetric.  Note first that if $\Bcs_1$ is omnigenous but not quasi-symmetric, then $\widetilde{\alpha}$ must be a non-constant function of $\varphi$.  However, the function $1/\cos(\widetilde{\alpha})$ must then reach a global maximum at some value of $\varphi$.  The coordinate $\zeta$ can be independently optimized to maximize the function $\cos(\zeta - \widetilde{\alpha})$, \ie $\zeta = \widetilde{\alpha}$ or $\zeta = \widetilde{\alpha} + \pi$, depending on the desired sign.  This means that the magnetic field reaches its maximum at a point in the $\zeta$-$\varphi$ plane, which is inconsistent with the required topology of contours of constant $\eta$.  Therefore we reach a contradiction and conclude that $\widetilde{\alpha}$ must be a constant, \ie the only possible omnigenous fields with toroidally or helically closed contours of magnetic field strength are quasi-symmetric.

\subsubsection{Poloidally closed lines of magnetic field strength}

Finally, we consider the case of poloidally closed lines of constant field strength, taking $\zeta = \varphi$ and $\xi = \theta$.  We can show that pseudo-symmetric fields are altogether impossible.  Fixing $\eta = \eta_c$ such that $\Bcs_1(\eta_c) \neq 0$, we consider a poloidal transit $\theta \rightarrow \theta + 2\pi$.  While the lefthand side of Eqn.~\ref{CS-B1-const} remains constant, the argument of $\cos$ on the righthand side must increase by $2\pi$ since the $\varphi$ returns to its initial value at the end of the transit.  Therefore, by the intermediate value theorem, this argument must pass through a zero of $\cos$, making the righthand side zero, contradicting the assumption that $\Bcs_1 \neq 0$.  Thus we find the only fields that satisfy Eqn.~\ref{CS-B1-const} are $\Bcs_1 = 0$.

%

\subsection{The case of varying on-axis magnetic field strength (quasi-isodynamic fields)}
\label{sec:quasi-isodynamic}

We now consider the on-axis magnetic field strength to vary with $\varphi$ (\ie $\partial \Ba/\partial \varphi \neq 0$ almost everywhere).  This restricts us to the class of poloidally closed magnetic fields (\eg quasi-poloidal symmetric and quasi-isodynamic fields).  We thus henceforth fix the angles of the CS mapping to be $\zeta = \varphi$ and $\xi = \theta$.  For this case we find it is possible to construct first-order omnigenous fields, at least in a certain approximate sense.  Note that quasi-poloidal symmetry cannot be achieved at first order because it requires $d$, a quantity proportional to the axis curvature, to be zero.  A magnetic axis with zero curvature everywhere cannot be closed.

We now set the two forms of the magnetic field strength equal, \ie by Eqns.~\ref{om-B} and \ref{GB-B}, and use $\eta_0 = \varphi$ to obtain at zeroth order in $\epsilon$

\begin{equation}
\Ba(\varphi) = \Bcs_0(\varphi),\label{B0-eqn}
\end{equation}
and, at first order,

\begin{equation}
\Ba(\varphi) d \cos(\theta - \alpha) = \eta_1 \Bcs_0^{\prime}(\varphi) + \Bcs_1(\varphi).\label{B1-eqn}
\end{equation}
The first thing we notice is that although the lefthand side depends on $\theta$, the term $\Bcs_1$ on the righthand side depends only on $\varphi$, therefore, it must balance and cancel with part of the other term on the righthand side.  Formally, we may filter out the $\theta$-averaged part and ignore it since, by \ref{B1-eqn}, it cannot affect the magnetic field strength, so it represents non-physical freedom in the function $\Bcs(\eta)$.  That is, we assume $\eta_1$ to henceforth have no $\theta$-average, and set $\Bcs_1 = 0$ without loss of generality.  (This is entirely consistent with Eqns.~\ref{F1-CS}-\ref{eta-1-eqn} since the filtering commutes with the $\zeta$-shift).

Solving Eqn.~\ref{B1-eqn} for $\eta_1$ and using Eqn.~\ref{eta-1-eqn}, we obtain

\begin{equation}
F_1(\theta, \varphi) = -\frac{\Bcs_0(\varphi) d(\varphi)}{\Bcs_0^\prime(\varphi)} \cos(\theta - \alpha(\varphi)).\label{F1-GB-eqn}
\end{equation}
Now we observe that the omnigenous form of $F_1$, Eqn.~\ref{F1-CS}, expresses a sort of symmetry that can be written as 

\begin{equation}
F_1(\theta, \eta) = F_1(\theta - \iotaslash \Delta\eta(\eta), \eta_b(\eta)),\quad\text{for }\eta\in D_{iR}.
\end{equation}
Applying this condition to the expression for $F_1$ given in Eqn.~\ref{F1-GB-eqn}, we obtain an equation of the form $a \cos(\theta + t_1) = b \cos(\theta + t_2)$ from which it follows that $a = \pm b$ and $t_1 = t_2 + 2\pi n + (\pi \mp \pi)/2$; we choose the upper sign convention and take $n = 0$, obtaining

\begin{equation}
d(\varphi) = \eta_b^\prime(\varphi) d(\eta_b(\varphi)),\quad\text{for }\varphi\in D_{iR},\label{d-constraint}
\end{equation}
and

\begin{equation}
\alpha(\varphi) = \alpha(\eta_b(\varphi)) + \iotaslash \Delta\eta(\varphi),\quad\text{for }\varphi\in D_{iR}.\label{alpha-constraint}
\end{equation}
To obtain Eqn.~\ref{d-constraint}, note that $\Bcs_0(\eta_b(\varphi)) = \Bcs_0(\varphi)$ implies $\Bcs_0^\prime(\varphi) = \eta_b^\prime(\varphi) \Bcs_0^\prime (\eta_b(\varphi))$.  Eqn.~\ref{alpha-constraint} can be thought of as providing a way to construct the function $\alpha(\varphi)$ for $\varphi \in D_{iR}$, given its dependence within $D_{iL}$.  Eqn.~\ref{d-constraint} provides a similar prescription for $d(\varphi)$.  Note that, assuming irrational $\iotaslash$, the function $\alpha$ cannot be consistently defined to force periodicity.  That is, evaluating Eqn.~\ref{alpha-constraint} at the global maximum, $\varphi = 2\pi$, we obtain $\alpha(2\pi)-\alpha(0) = 2\pi \iotaslash$, which can only be a multiple of $2 \pi$ if $\iotaslash$ is an integer.  This does not actually affect the periodicity of the first order magnetic field, since, from Eqn.~\ref{B1-eqn} and $\Bcs_0^\prime(0) = 0$, we must have $d(0) = 0$, implying that it is zero at $\varphi = 0$ (and therefore is periodic).  In fact this must be true at all extrema of $\Bcs_0$,

\begin{equation}
d = 0,\quad\text{at all local extrema,}\label{d-constraint-2}
\end{equation}
which also follows from Eqn.~\ref{d-constraint} if $d$ is to be continuous at these locations.  Even if the field strength is periodic, this does not automatically imply periodicity in derivatives of the magnetic field strength, but this would be too much to hope for, given that omnigenous solutions are generally non-analytic \citep{cary-shasharina}.  As we will see, periodicity is also not be so easily enforced for the full solution, but in the next section we will propose a way to repair these discontinuities.

\section{Constructing quasi-isodynamic equilibria}\label{sec:QI-theoretical-construction}

In the previous section, we have demonstrated that the first order magnetic field strength of the near-axis expansion can be made consistent with the condition of omnigenity.  This alone is not sufficient to solve the equilibrium problem.  We must also demonstrate that there is a consistent solution of the equilibrium equation introduced by \citet{garren-boozer-1}.

To apply the results of \citet{garren-boozer-1} to the present problem, however, we must account for the fact that the curvature of the magnetic axis will be zero at some toroidal locations (this follows from Eqn.~\ref{d-constraint-2} as shown below).  The conventional Frenet frame is discontinuous at points of zero curvature, when the normal and binormal unit vectors change their signs, which can be considered the typical case since it corresponds to first order zeros of the curvature.  Fortunately, we can employ a modified Frenet frame that accounts for these sign flips; a formal development of such frames is given by \citet{Carroll_2013}.  Under reasonable conditions, this frame will be continuous everywhere, and its derivatives satisfy precisely the same equations as the traditional Frenet frame, so that the derivations of \citet{garren-boozer-1} proceed identically, with only the trivial modification of replacing the curvature with the ``signed curvature'' $\kappa^s(\varphi)= s(\varphi) \kappa(\varphi)$, and reinterpreting the components of the coordinate mapping using the modified frame.  Note that the sign $s(\varphi)$ takes values $+1$ or $-1$, and switches at locations of zero curvature, and the normal and binormal unit vectors of the modified Frenet frame are given by multiplying those of the traditional Frenet frame by $s(\varphi)$.

The equation that must be solved to find an equilibrium solution is a first order nonlinear ODE for the quantity $\sigma(\varphi)$, which can be interpreted as follows with the modified Frenet frame: the coordinate mapping is expressed to first order as ${\boldsymbol x} \approx {\boldsymbol r}_0 + X_1 {\boldsymbol n}^s + Y_1 {\boldsymbol t}^s$, where ${\boldsymbol r}_0(\varphi)$ specifies the magnetic axis, and, via its derivatives, also the signed normal and binormal unit vectors, ${\boldsymbol n}^s(\varphi)$ and ${\boldsymbol t}^s(\varphi)$.  The iterative solution of the equilibrium equations leads to an expressions for the components $X_1$ and $Y_1$, so that the quantity $\sigma$ is seen to be related to the amplitude of the in-phase part of the binormal component of the coordinate mapping (see \cite{garren-boozer-1, landreman-sengupta} for more details),

\begin{eqnarray}
X_1 = \bar{d}(\varphi) \cos[\theta - \alpha(\varphi)],\label{X1-eqn}\\
Y_1 = \frac{2}{\bar{d}(\varphi)} \{  \sin[\theta - \alpha(\varphi)] + \sigma(\varphi) \cos[\theta - \alpha(\varphi)] \},\label{Y1-eqn}
\end{eqnarray}
where $\bar{d}$ is related to $d$ as

\begin{equation}
d(\varphi) = \bar{d}(\varphi) \kappa^s(\varphi).
\end{equation}
Note that from this and Eqn.~\ref{d-constraint-2} we see that the curvature must indeed be zero at extrema of $B_a$.  Although $d$ generally changes sign at points of zero curvature, $\bar{d}$ can be continuous and non-zero at these places.  Note that $\bar{d}$ cannot have any zeros, for $Y_1$ to remain finite.  Because of this, $\kappa^s$ and $d$ must have zeros of the same order.  Furthermore, the periodicity of $X_1$ cannot be be achieved by setting the coefficient, $\bar{d}$, of Eqn.~\ref{X1-eqn} to zero.  Instead, periodicity of $X_1$ and $Y_1$ implies the constraints



\begin{eqnarray}
\bar{d}(2\pi) = \bar{d}(0),\\
\alpha(2\pi) - \alpha(0) = 2\pi N \text{, for some integer }N,\label{alpha-periodicity}\\
\sigma(2\pi) = \sigma(0),\label{sigma-periodicity}
\end{eqnarray}
where the first should already be satisfied if $d$, satisfies Eqn.~\ref{d-constraint}.  Given irrational $\iotaslash$, the second of these conditions is not consistent with omnigenity (Eqn.~\ref{alpha-constraint}), which implies $\alpha(2\pi) - \alpha(0) = 2\pi \iotaslash$.  We will propose a possible way to overcome this in the next section.

Finally, the equation that must be solved for $\sigma(\varphi)$ is
\begin{equation}
\sigma^\prime + (\iotaslash - \alpha^\prime)(\sigma^2 + P) + Q = 0.\label{sigma-eqn}
\end{equation}
where $P = 1 + \bar{d}^4/4$, $Q = -G_0\bar{d}^2(\tau + I_2/2)$, with $G_0$, $I_2$ and $\tau(\varphi)$ being related to the poloidal current, toroidal current and torsion of the magnetic axis, respectively.  

Let us now summarize the problem of finding an omnigenous magnetic field at first order: first one can freely specify the on-axis magnetic field strength $\Bcs_0(\varphi)$, an axis shape, (having zero curvature at the extrema of $\Bcs_0$), and a function $d(\varphi)$ satisfying Eqn.~\ref{d-constraint}.  A function $\alpha(\varphi)$ can also be arbitrarily specified on $D_L$ but its dependence on $D_R$ is constrained by Eqn.~\ref{alpha-constraint} which depends on $\iotaslash$.  Therefore the solution of Eqn.~\ref{sigma-eqn} for $\sigma(\varphi)$ and $\iotaslash$, must be done consistently; see Sec.~\ref{sec:numerical-solutions} for the demonstration of this.  In principle, this would complete the solution.  However, as we have discussed, there is a conflict between Eqn.~\ref{alpha-constraint} and the periodicity constraint on $\alpha$, Eqn.~\ref{alpha-periodicity}, and in the following section we propose a practical way to address this.

\subsection{Controlled approximation of omnigenous fields}
\label{sec:approximately_omnigenous}

Because an omnigenous magnetic field (that is not quasi-symmetric) is necessarily non-analytic, such a field can only ultimately be physically realized by a smooth approximation.  One approach, proposed by \citet{cary-shasharina}, is to truncate a series representation of an exactly omnigenous field, resulting in a smooth but slightly non-omnigenous one.  Another idea is to specify a sufficiently smooth magnetic field that violates omnigenity in a controlled way -- this, as we show below, can also simultaneously help resolve the conflict between omnigenity and periodicity, which was encountered in the previous section.

To ensure the appropriate behavior of $\alpha(\varphi)$, Eqn.~\ref{alpha-periodicity}, we introduce small matching regions near $\varphi = 0, 2\pi$, where $\alpha$ will be defined such that condition \ref{alpha-periodicity} is satisfied, while abandoning the condition of omnigenity there:

\begin{align}
\alpha(\varphi) = \begin{cases}
\alpha_\mathrm{I}(\varphi),&\text{ for }0 \leq \varphi \leq \delta,\\
\alpha_\mathrm{II}(\varphi),&\text{ for }\delta \leq \varphi \leq 2\pi - \delta,\\
\alpha_\mathrm{III}(\varphi),&\text{ for }2\pi - \delta \leq \varphi \leq 2\pi,
\end{cases}
\end{align}
In region II where omnigenity is satisfied, Eqn.~\ref{alpha-constraint} implies that $\alpha_\mathrm{II}$ may be written in terms of the function $a(\varphi)$, defined on $D_L$:

\begin{align}
\alpha_\mathrm{II}(\varphi) = \begin{cases}
a(\varphi),&\text{ for } \varphi \in D_{iL}\\
a(\varphi_b(\varphi)) + \iotaslash \Delta\varphi(\varphi),&\text{ for } \varphi \in D_{iR}.
\end{cases}
\end{align}
Introducing $c(\varphi) = a(\varphi) + (\iotaslash/2)\Delta\varphi(\varphi_b(\varphi))$, extending the definition of $\eta_b(\varphi)$ to $D_L$ such that it is its own inverse, $\eta_b = \eta_b^{-1}$, we can write $\alpha$ in a more symmetric form:

\begin{align}
\alpha_\mathrm{II}(\varphi) = \begin{cases}
c(\varphi) + (\iotaslash/2)\Delta\varphi(\varphi),&\text{ for } \varphi \in D_{iL}\\
c(\varphi_b(\varphi)) + (\iotaslash/2)\Delta\varphi(\varphi),&\text{ for } \varphi \in D_{iR},
\end{cases}\label{alpha-2-eqn-2}
\end{align}
where we have used that $\Delta\varphi(\varphi) = \varphi_b(\varphi) - \varphi$, so $\Delta\varphi(\varphi_b(\varphi)) =  -\Delta\varphi(\varphi)$.

To make further progress, let us make some simplifying assumptions.  We assume symmetric, single-well $\Bcs_0(\varphi)$, so that $\varphi_b(\varphi) = 2\pi - \varphi$, and $\Delta\varphi(\varphi) = 2\varphi - 2\pi$.  Eqn.~\ref{sigma-eqn} can be written as

\begin{equation}
\sigma^\prime + \gamma(\varphi)(\sigma^2 + P) + Q = 0,\label{sigma-eqn-2}
\end{equation}
where $\gamma(\varphi) = \iotaslash - \alpha^\prime$.  The functions $\alpha_\mathrm{I}$ and $\alpha_\mathrm{III}$ must be chosen such that $\alpha(2\pi)-\alpha(0) = 2\pi N$, but are otherwise free.  However, a particularly simple prescription is found as follows.  We note that $\alpha_\mathrm{II}$, as expressed in Eqn.~\ref{alpha-2-eqn-2}, is composed of a secular piece ($\iotaslash/2\Delta\varphi(\varphi)$), and an even period piece, which need not be corrected to satisfy periodicity.  Thus, we extend $\alpha_\mathrm{II}$ into the buffer regions, and add a correction to the secular part.  That is, we write $\alpha_\mathrm{I} = \alpha_\mathrm{II} + \Delta\alpha_\mathrm{I}$ and $\alpha_\mathrm{III} = \alpha_\mathrm{II} + \Delta\alpha_\mathrm{III}$, where continuity of $\alpha$ requires that the correction goes to zero at the interface with the buffer ($\Delta\alpha_\mathrm{I}(\delta) = 0$, and $\Delta\alpha_\mathrm{III}(2\pi-\delta) = 0$), and the periodicity constraint is satisfied via $\alpha_\mathrm{I}(0) = -N\pi$ and $\alpha_\mathrm{III}(2\pi) = N\pi$.  Taking $\Delta\alpha_\mathrm{I}(\varphi)$ and $\Delta\alpha_\mathrm{III}(\varphi)$ to be linear functions of $\varphi$ then uniquely determines them, resulting in

\begin{align}
    \alpha = \alpha_\mathrm{II} + \pi(N - \iotaslash) 
    \begin{cases}
        \varphi/\delta - 1,&\text{ for }0 \leq \varphi < \delta,\\
        0,&\text{ for }\delta \leq \varphi \leq 2\pi-\delta,\\
        (\varphi - 2\pi)/\delta + 1,&\text{ for }2\pi-\delta < \varphi \leq 2\pi.
    \end{cases}
\end{align}
The function $\gamma$ is then simply

\begin{align}
\gamma(\varphi) = \gamma_\mathrm{o}(\varphi) + \begin{cases}
(\iotaslash-N)\pi/\delta,&\text{ for }0 \leq \varphi \leq \delta,\text{ or }2\pi - \delta \leq \varphi \leq 2\pi,\\
0,&\text{ for }\delta \leq \varphi \leq 2\pi - \delta,\\
\end{cases}
\end{align}
where $\gamma_\mathrm{o}(\varphi)$ is the odd function

\begin{align}
\gamma_\mathrm{o}(\varphi) = \begin{cases}
-c^\prime(\varphi),&\text{ for } \eta \in [0,\pi]\\
c^\prime(2\pi - \varphi),&\text{ for } \eta \in [\pi, 2\pi].
\end{cases}
\end{align}

We note that the integer $N$, introduced in Eqn.~\ref{alpha-periodicity}, must be consistently set, taking into account the number of times the coordinate mapping rotates around the magnetic axis during a toroidal transit, so that $\theta$ behaves as a proper poloidal angle -- \ie it increases by $2\pi$ with a poloidal transit, but is periodic in the toroidal direction.

Note that $\iotaslash$ only appears in the definition of $\gamma(\varphi)$ in the matching regions, so we may consider solving Eqn.~\ref{sigma-eqn-2} independently in the omnigenous region.  Then, $\iotaslash$ can be tuned in the matching regions to achieve $2\pi$ periodicity of $\sigma$.  Assuming that, for small $\delta$, the functions $P$, $Q$, and $\gamma$ can be considered constant in the matching regions, the proof of the existence and uniqueness of this value of $\iotaslash$ in fact follows easily from the results of \cite{landreman-sengupta-plunk}.  A general existence and uniqueness theorem, \ie allowing general $\alpha_\mathrm{I}$, $\alpha_\mathrm{III}$ and $\Bcs_0$, appears more difficult to prove.


\section{Numerical solutions}\label{sec:numerical-solutions}

\subsection{Algorithm}

We now describe how the equations of section \ref{sec:quasi-isodynamic} can be solved in practice numerically.
The inputs to the calculation are the shape of the magnetic axis, the functions $B_0 = \Bcs_0(\varphi)$ and $d(\varphi)$, a chosen width for the buffer regions $\alpha_{\mathrm{I}}$ and $\alpha_{\mathrm{III}}$, and a finite value of aspect ratio.
The outputs of the calculation are $\iotaslash$, $B_1(\theta,\varphi) = B_0(\varphi) d \cos(\theta - \alpha)$, and the shape of the elliptical flux surfaces.

The first step in the numerical solution is to compute $G_0$ and $\varphi(\phi)$ along the axis, where $G_0$ is the on-axis value of the coefficient $G$ in ${\bf B}_{\mathrm{cov}}$, and $\phi$ is the standard toroidal angle (i.e. the azimuthal angle for cylindrical coordinates).
These calculations can be done by iteratively solving (2.20)-(2.22)
in \cite{landreman-sengupta},
\begin{align}
\frac{d\varphi}{d\phi}= \frac{\ell^' B_0}{|G_0|},
\hspace{0.5in}
|G_0| = \frac{1}{2\pi}\int_0^{2\pi}d\phi \;B_0(\varphi(\phi)) \ell^',
\end{align}
where $\ell^' = |d \vect{r}/d \phi|$ is the arclength increment. Beginning with the guess $\varphi \approx \phi$, fixed-point (Picard) iteration converges quickly.
The sign of $G_0$ (i.e. whether ${\bf B}$ points towards increasing or decreasing $\varphi$) is a free input.

Next we solve the equation for $\sigma$, (\ref{sigma-eqn}), using Newton's method.  This solution procedure is similar to the one for quasisymmetry described in section
3 of \citep{landreman-sengupta-plunk}, but with a few modifications.  The discrete unknowns include values of $\sigma$ on a uniform grid of $N_\phi$ points in the domain $\phi \in [0,2\pi/n_{fp}]$,
where $n_{fp}$ is the number of identical field periods.
The $d\sigma/d\varphi$ term in (\ref{sigma-eqn}) is discretized using the pseudospectral differentiation matrix.
For computing omnigenous solutions, it is convenient if none of the grid points lie exactly where the curvature vanishes; otherwise
(\ref{sigma-eqn}) would require evaluating $d/\kappa=0/0$. To avoid points of vanishing curvature at both $\phi=0$ and $\phi = \pi/n_{fp}$ (half period), the grid points are shifted by one third of the grid spacing: $\phi_j = (j-2/3)2\pi/n_{fp}$ for $j=1 \ldots N_\phi$.

In addition to the unknowns $\sigma(\phi_j)$ there is one more unknown: $\iotaslash$. Corresponding to this additional unknown is one additional equation, reflecting the freedom to specify $\sigma(0)$ (as discussed in the Appendix of \cite{landreman-sengupta-plunk}.) In the common case of stellarator symmetry, this extra condition is $\sigma(0)=0$. While $\varphi=0$ is not one of the grid points, this condition can nonetheless be imposed by interpolating $\sigma$ from the grid points $\phi_j$ (using pseudospectral interpolation.)

To apply Newton's method, the Jacobian is needed. One block of the Jacobian corresponds to the derivative of (\ref{sigma-eqn}) with respect to $\sigma$, which is straightforward. Another block corresponds to the derivative of (\ref{sigma-eqn}) with respect to $\iotaslash$, and in contrast to the quasisymmetric case, here we must account for the fact that $\alpha$ depends on $\iotaslash$. Fortunately, in both the central region and buffer regions, $\alpha$ depends on $\iotaslash$ as 
\begin{align}
    \alpha(\varphi,\iotaslash)\ = \iotaslash\alpha_{\iotaslash}(\varphi) + \alpha_{0}(\varphi),
    \label{eq:alpha_dependence_on_iota}
\end{align}
for some functions $\alpha_{\iotaslash}(\varphi)$ and $\alpha_{0}(\varphi)$. Therefore the Jacobian block
corresponding to the derivative of (\ref{sigma-eqn}) with respect to $\iotaslash$ is given by
$(1-\alpha^{\prime}_{\iotaslash})(\sigma^2+P)$.

Finally, once 
$\sigma$ with self-consistent $\iotaslash$
has been found, the result can be converted to cylindrical coordinates and a finite-aspect-ratio configuration can be generated using any of the methods described in Section 4 of \cite{landreman-sengupta-plunk}.  For results shown below, we use the method of section 4.1.  The resulting toroidal boundary shape in cylindrical coordinates can then be supplied to an equilibrium code such as VMEC \citep{Hirshman}.

Instead of solving equation (\ref{sigma-eqn}) for $\sigma$, it is also possible to directly solve the equations for the near-axis flux surface shape in cylindrical coordinates derived in \cite{landreman-sengupta}. This approach is conceptually valuable since it makes no use of the Frenet frame, so one need not worry about discontinuities of the Frenet representation. The numerical solution can be obtained using Newton's method, as detailed in section 4.3 of \cite{landreman-sengupta-plunk}. In contrast to the quasisymmetric case, for the omnigenous case $B_1$ depends on $\iotaslash$ through $\alpha$. The block of the Jacobian corresponding to the derivative of the residual with respect to $\iotaslash$ can again be evaluated by noting (\ref{eq:alpha_dependence_on_iota}).
We have implemented both this cylindrical coordinate approach and the Frenet ($\sigma$) approach, and verified the results from the two methods converge towards each other as $N_\phi \to \infty$.
For instance, in the example of the next section with $N_\phi=101$ grid points, the two approaches yield the same value of $\iotaslash$ through 9 digits.


\subsection{Example}
\label{sec:example}

\begin{figure}
\centering
\includegraphics[width=2.4in]{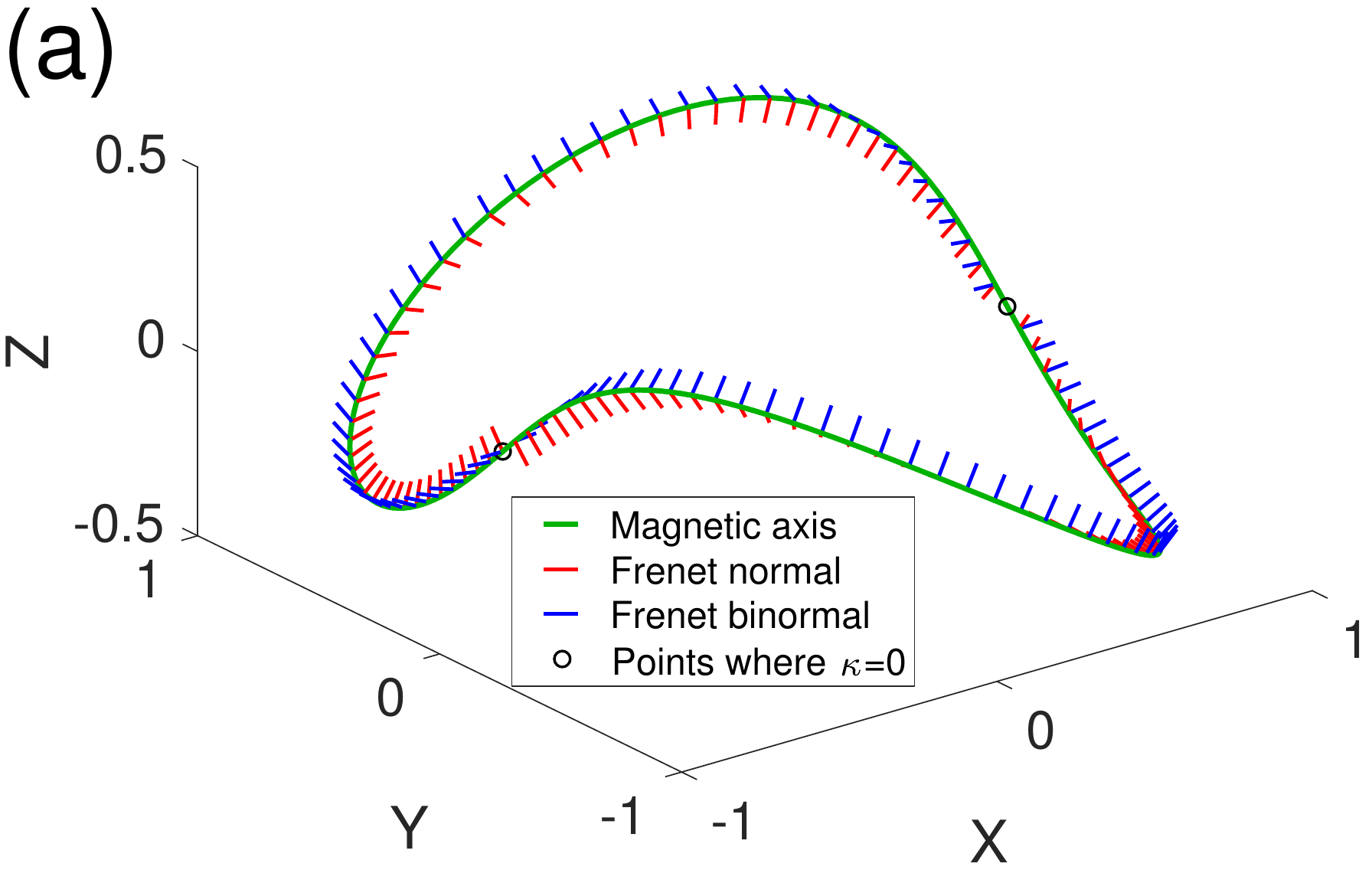}
\includegraphics[width=2.4in]{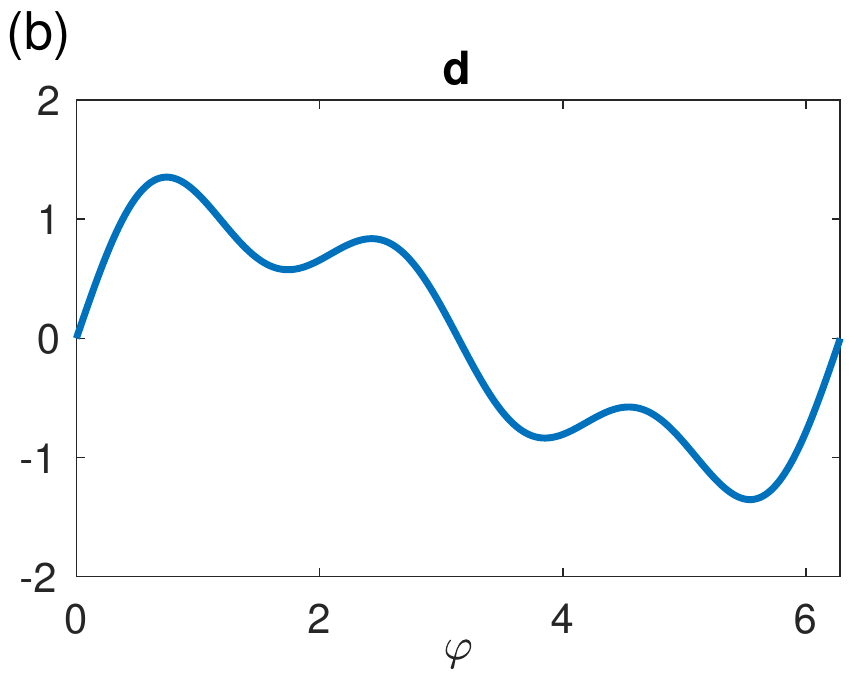}
\includegraphics[width=2.4in]{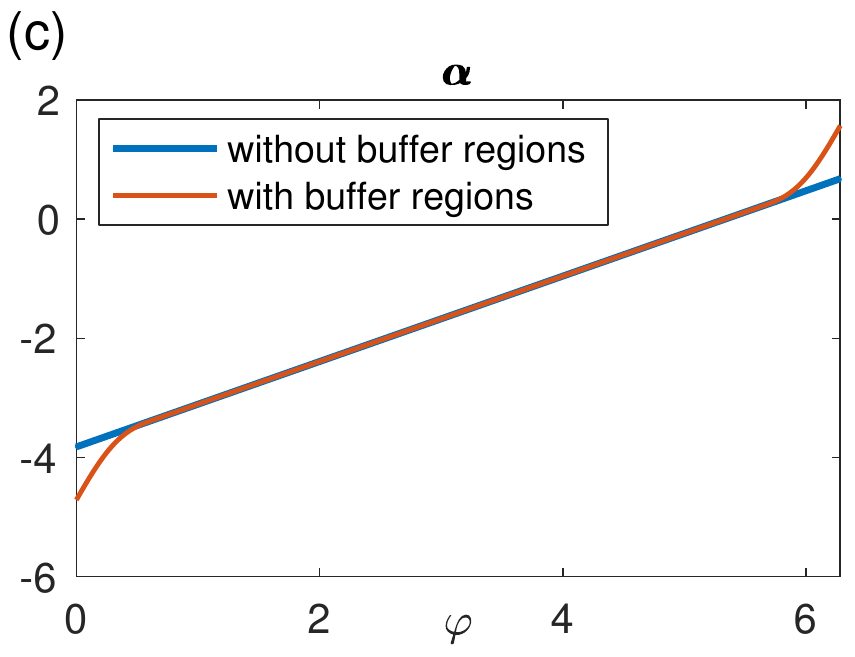}
\includegraphics[width=2.4in]{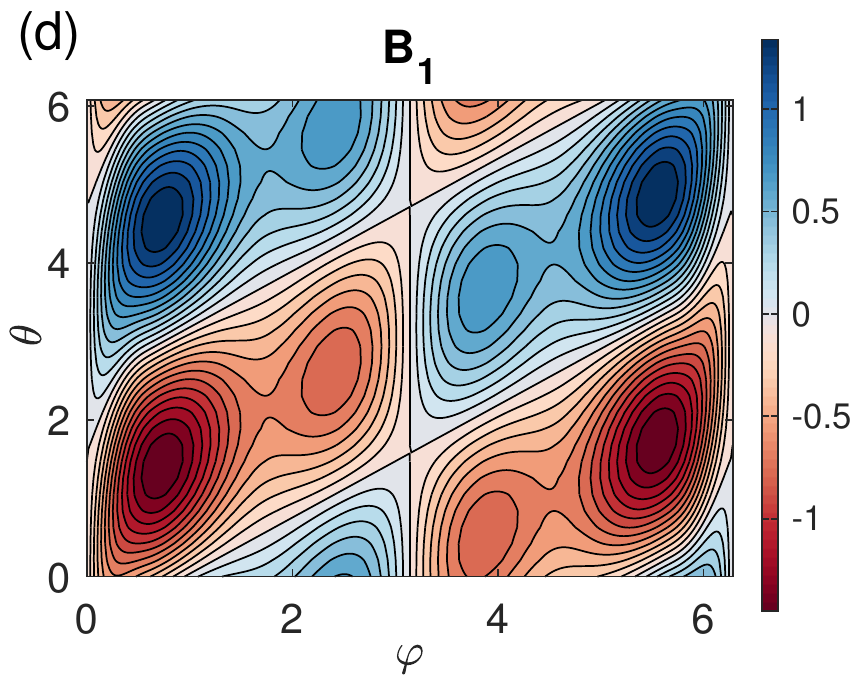}
\caption{Elements of the numerical construction include (a) the shape of the magnetic axis, (b) the function $d(\varphi)$, (c) the phase $\alpha(\varphi)$, and (d) the first order (in $\epsilon$) variation of the field strength, $B_1$.}
\label{fig:numerical_solution_components}
\end{figure}
\begin{figure}
\centering
\includegraphics[width=2.4in,valign=t]{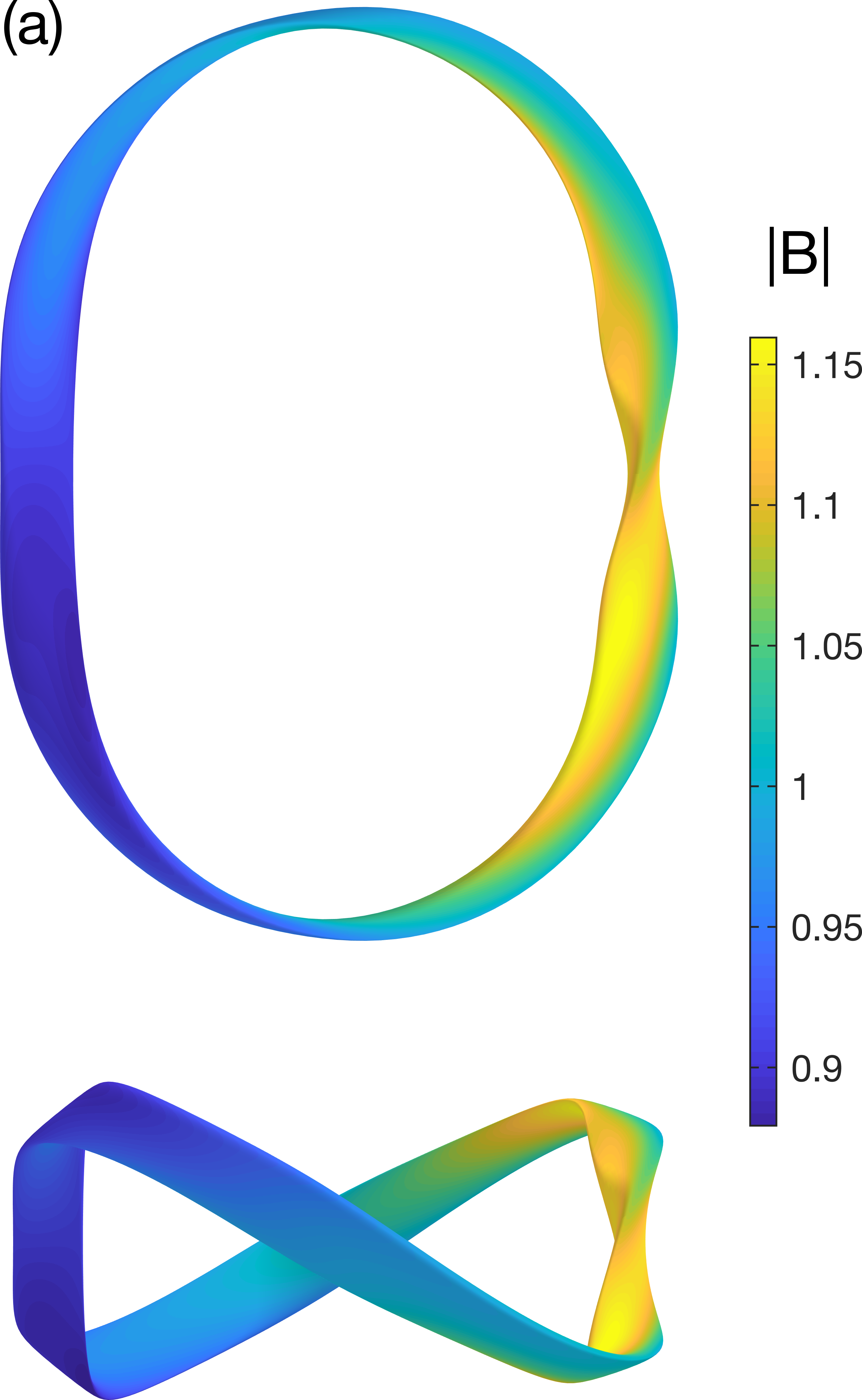}
\includegraphics[width=2.4in,valign=t]{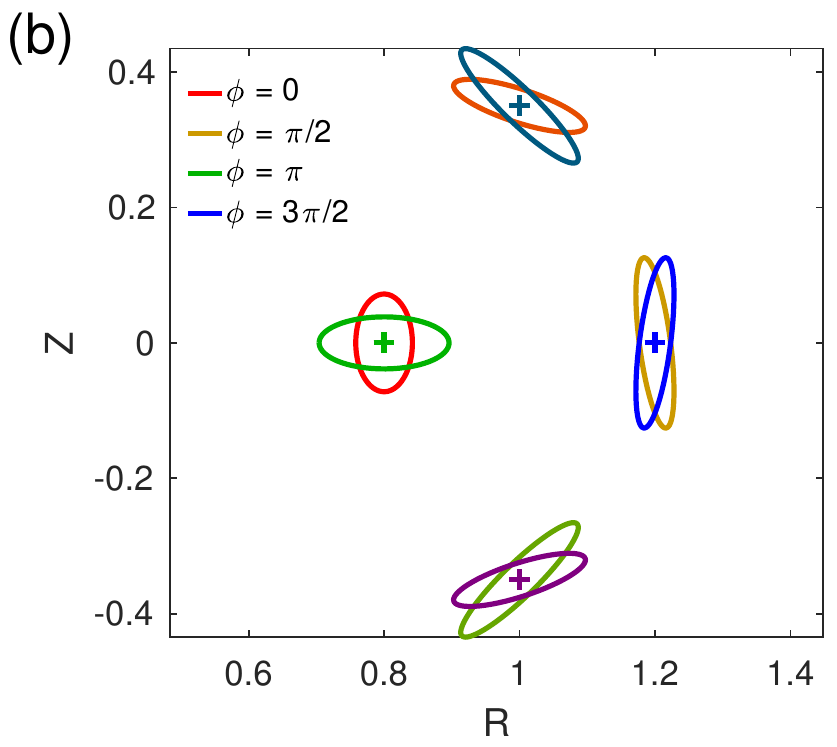}
\caption{The shape of the constructed omnigenous configuration, for aspect ratio 20. (a) Boundary shape in 3D, with color indicating $B$ computed by VMEC. A bird's eye view and side view are shown. (b) Cross-sections of the configuration at 8 values of toroidal angle $\phi$.}
\label{fig:numerical_solution_shape}
\end{figure}
\begin{figure}
\centering
\includegraphics[width=2.5in]{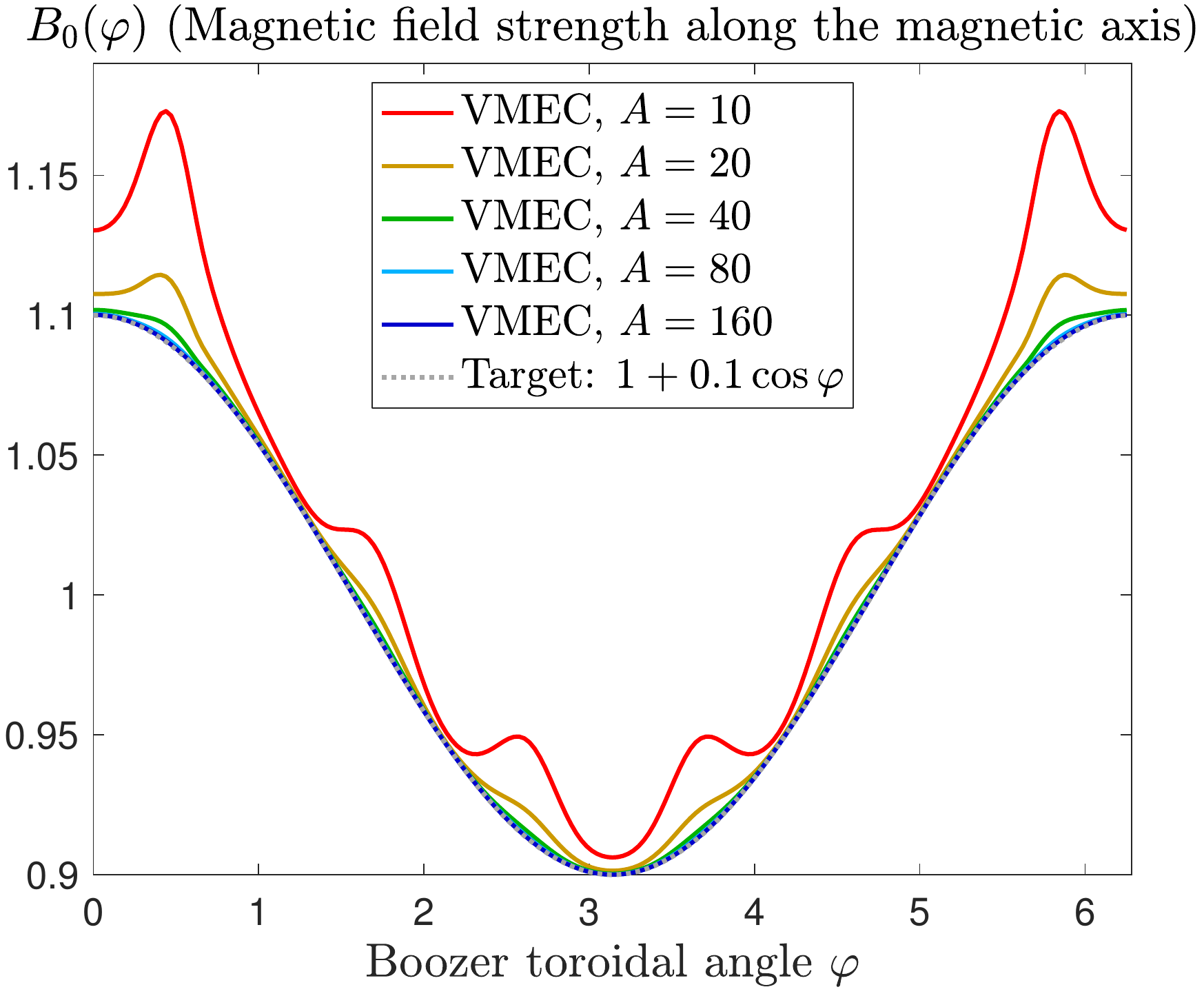}
\caption{As the aspect ratio $A$ used for the construction increases, $B$ on the axis of the numerical VMEC equilibrium inside the constructed boundary (solid colored curves) converges to the desired target function (dotted gray curve).
}
\label{fig:B0_convergence}
\end{figure}
\begin{figure}
\centering
\includegraphics[width=5in]{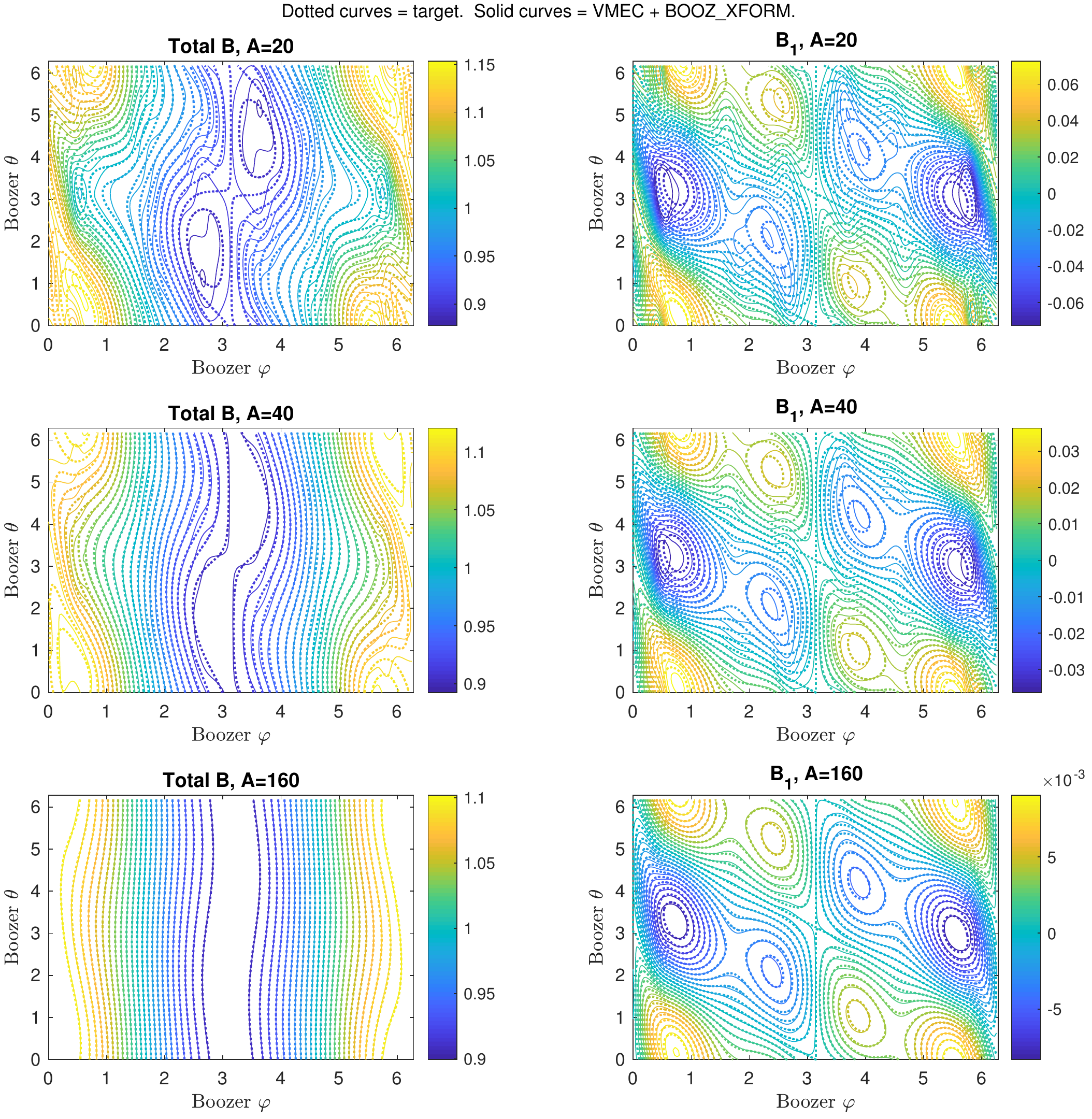}
\caption{As the aspect ratio $A$ used for the construction increases, $B$ for the numerical VMEC equilibrium inside the constructed boundary (solid curves) converges to the desired target function (dotted curves). Left column shows the total $B$ at the boundary; right column shows $B - (1+0.1\cos\varphi)$ at the boundary.
}
\label{fig:B1_convergence}
\end{figure}

An example of an omnigenous non-quasisymmetric
configuration constructed using this procedure
is shown in figures \ref{fig:numerical_solution_components}-\ref{fig:B1_convergence}. We begin by choosing the
axis shape
\begin{align}
R_0(\phi)=1 - 0.2 \cos(2\phi),
\hspace{0.5in}
z_0(\phi)=0.35 \sin(2\phi),
\end{align}
shown in figure \ref{fig:numerical_solution_components}.a, which has vanishing curvature at two points:
$\phi=0$ and $\pi$.
The Frenet normal and binormal vectors both display a discontinuous reversal in direction at these points.
When viewed from above, \eg from a viewpoint with $R=0$ and $z > 0$, the axis has the shape of a racetrack oval, with the points of vanishing curvature at the middle of each straightaway. The oval is twisted out of the $z=0$ plane so it has net torsion.
We choose the on-axis field strength to be
$B_0(\varphi) = 1 + 0.1 \cos\varphi$.
Note that while the axis shape is symmetric under rotation by $\pi$, as in a device with two field periods, the field
strength does not have this symmetry, so the number of field periods is one.

The elongation of the final configuration is sensitive to the choice of $d(\varphi)$. It can be challenging to find a function $d(\varphi)$ for which the elongation does not grow to impractically large values, $\gg 10$. For the configuration in the figures we use 
\begin{align}
d(\varphi) = 1.08 \sin(\varphi) + 0.26 \sin(2\varphi) + 0.46 \sin(3 \varphi),
\end{align}
where the coefficients were chosen to keep the elongation down at tolerable levels, $\sim 5$. Further tailoring of the $d(\varphi)$ function might yield lower values of elongation.

We choose $\alpha=-3\pi/2$ at $\varphi=0$ and $\alpha=\pi/2$ at $\varphi=2\pi$. In region II, $\alpha = -\pi/2 + (\varphi - \pi)\iotaslash$. We allocate the first 10\% and last 10\% of the toroidal domain to the buffer regions:  $\alpha_{\mathrm{I}}$  occupies $\varphi \in [0,\;\pi/5]$ and $\alpha_{\mathrm{III}}$ occupies $\varphi \in [9\pi/5,\; 2\pi]$.
In the buffer regions we choose $\alpha$ to be a fourth order polynomial in $\varphi$, enforcing continuity of $\alpha$ and its first two derivatives
at the domain boundaries $\varphi=0$, $\pi/5$, and $9\pi/5$.
(When $\alpha$ does not have a continuous derivative
at these boundaries, the VMEC solution exhibits numerical oscillations near these points.)

The shape of the resulting configuration is shown in figure
\ref{fig:numerical_solution_shape}, for aspect ratio 20.
It can be seen that the shape resembles a M\"{o}bius strip.
There is a pronounced twist in the shape near the region of maximum field strength.
The toroidal extent of this twist corresponds directly to the width of the buffer regions.
The constructed boundary shape is provided as input to the VMEC equilibrium code \citep{Hirshman}.
According to the near-axis construction, the rotational transform is 
$\iotaslash=0.717$. 
The rotational transform computed by VMEC is similar (0.70 on axis, 0.72 at the edge), and it converges
to the value predicted by the construction as the aspect ratio is increased.

\begin{figure}
\centering
\includegraphics[width=3in]{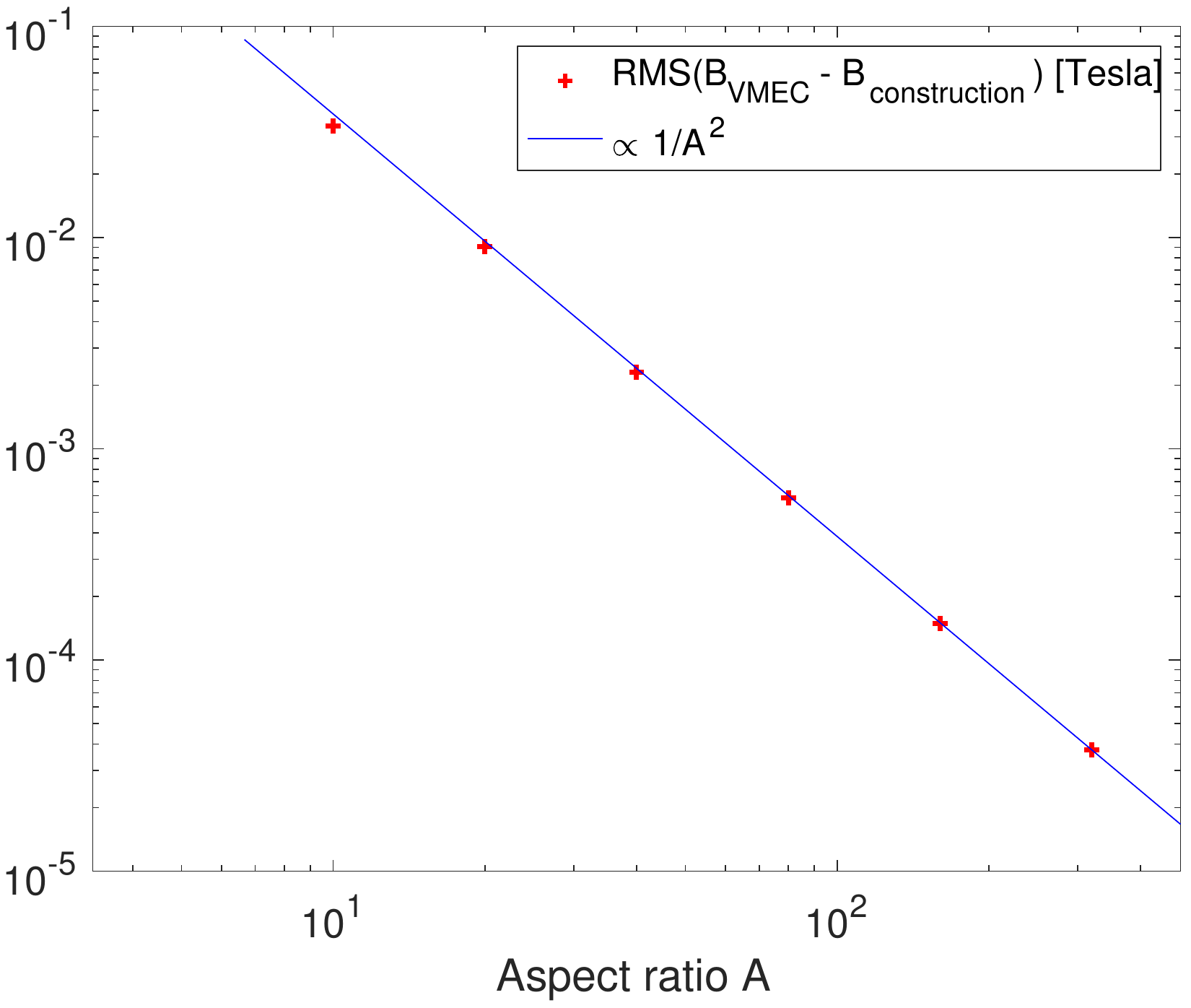}
\caption{The difference between $B$ predicted by the construction and $B$ computed by VMEC + BOOZ\_XFORM inside the constructed boundary, as measured by
the root-mean-square $\left[\int d\theta \int d\varphi \left( B_{\mathrm{VMEC}} - B_{\mathrm{construction}}\right)^2\right]^{1/2}$, scales as $A^{-2}$. This scaling is consistent with the fact that the construction here is carried out through $O(\epsilon)$.
}
\label{fig:B_A2_scaling}
\end{figure}

Given a VMEC solution, we then evaluate $B$ as a function of Boozer coordinates using the code BOOZ\_XFORM \citep{booz-xform}, to compare the achieved $B$ to the $B$ predicted by the near-axis construction.
We find that $B$ for the numerical equilibrium inside the constructed boundary (according to BOOZ\_XFORM) converges to the desired function $B(r,\theta,\varphi)$ as the aspect ratio $A$ increases. One aspect of this convergence can be seen in figure \ref{fig:B0_convergence}, which shows the convergence of the on-axis $B$ to the target function $1 + 0.1 \cos\varphi$. For $A \ge 80$, differences from the target function are barely discernible on the scale of the figure.  Furthermore, the convergence of $B$ at the boundary can be seen in figure \ref{fig:B1_convergence}. The total $B$ is shown in the left column, with the target $B_0$ subtracted in the right column, and the three rows show three increasing values of aspect ratio. As $A$ increases, the total $B$ converges to the desired function, and $B - B_0$ converges to the desired $B_1$.
Finally, figure \ref{fig:B_A2_scaling} shows how the difference between the target and achieved magnetic field strength scales with $A$. The difference between the target and achieved $B$ is measured by the root-mean-square (RMS) difference $\left[\int d\theta \int d\varphi \left( B_{\mathrm{VMEC}} - B_{\mathrm{construction}}\right)^2\right]^{1/2}$. In this formula, $B$ at the boundary is used.
This RMS difference is found to scale as $1/A^2$, as expected since the construction is done here through $O(\epsilon)$.

\subsection{Measuring deviation from omnigenity via $1/\nu$ transport}
\label{sec:eps_eff}

The collisionless confinement of trapped particle orbits also has a direct effect on neoclassical transport, which can be observed in the so-called $1/\nu$ transport, determined by the effective helical ripple, $\epsilon_{\mathrm{eff}}$ \citep{nemov-1999}:

\begin{equation}
    \epsilon_{\mathrm{eff}}^{3/2} = \left(\frac{\pi R^2}{2^{7/2}}\right) \lim_{L_s\rightarrow \infty} \left( \int_0^{L_s} \frac{dl}{B}\right)\left( \int_0^{L_s} \frac{dl|\bnabla \psi|}{B}\right)^{-2} \int_{B_\mathrm{min}/B_\mathrm{r}}^{B_\mathrm{max}/B_\mathrm{r}} dx \sum_j \frac{H_j^2(x)}{I_j(x)},\label{eps-eff}
\end{equation}
where the summation of the magnetic well index $j$ includes all wells in the interval $[0,L_s]$ containing points where $B = x B_\mathrm{r}$ is satisfied, $R$ is a reference major radius, $B_\mathrm{r}$ is a reference magnetic field strength, and

\begin{eqnarray}
    I_j = \sum_\gamma \gamma\int_{B_\mathrm{min}^j}^{x B_\mathrm{r}} \frac{dB}{B \partial B/\partial s}\sqrt{1-\frac{B}{B_\mathrm{r} x}},\\
    H_j = \frac{1}{x} \int_{B_\mathrm{min}^j}^{x B_\mathrm{r}} \frac{dB}{B^2}\sqrt{x - \frac{B}{B_\mathrm{r}}}\left(\frac{4B_\mathrm{r}}{B}-\frac{1}{x}\right) \sum_\gamma \gamma Y.
\end{eqnarray}
Here we again encounter the factor $Y$ of Eqn.~\ref{Y-def} that arises in the calculation of the bounce-averaged radial excursion $\Delta \psi$.  This expression is complicated, but we note that it is essentially the square of the average of $\sum_\gamma \gamma Y$.  For a perfectly omnigenous field, we therefore expect $\epsilon_{\mathrm{eff}} = 0$.  To test the quality of one of our constructed solutions, we generate an ``exact'' numerical solution by using the magnetic surface shape at a finite radial position as an input for the VMEC code.  Because we expect omnigenity to be satisfied only at first order (and only in the limit that the buffer regions are small), we expect $\sum_\gamma \gamma Y \sim {\cal O}(\epsilon^2)$.  Noting the factor of $|\bnabla\psi|^{-2} \sim {\cal O}(\epsilon^{-2})$ in Eqn.~\ref{eps-eff}, we find $\epsilon_{\mathrm{eff}}^{3/2} \sim {\cal O}(\epsilon^2)$.

\begin{figure}
\centering
\includegraphics[width=3in]{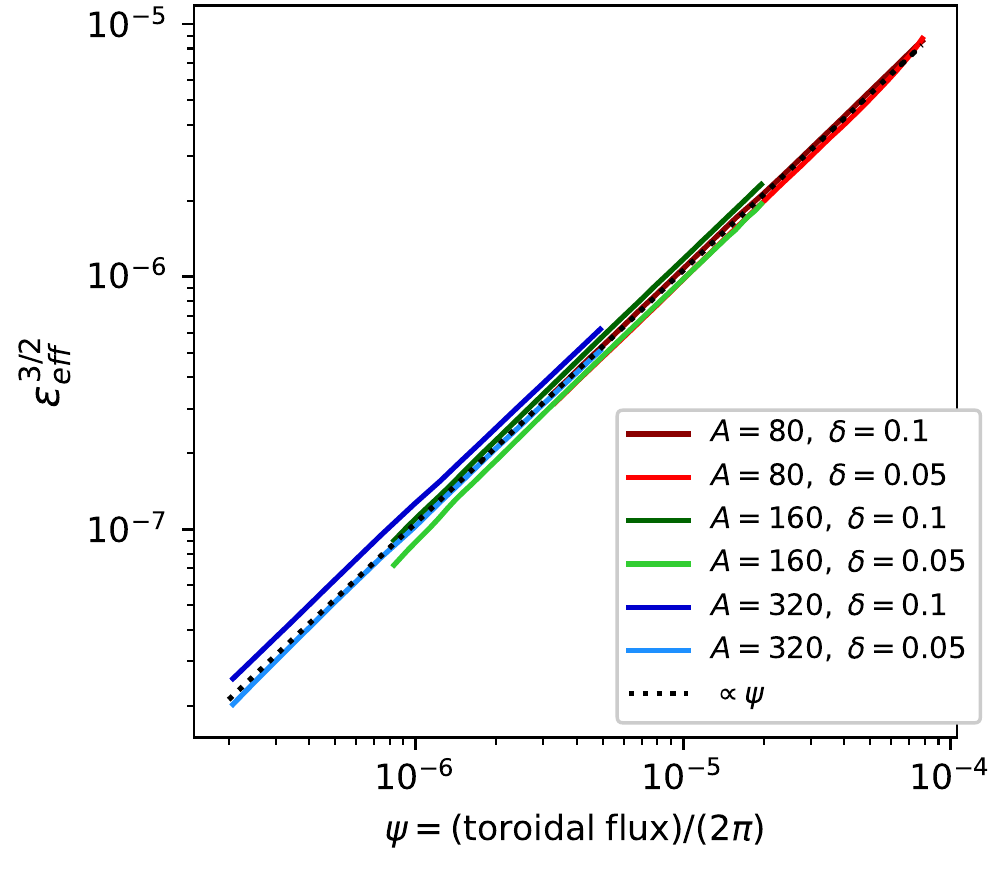}
\caption{For sufficiently large aspect ratio $A$ and small buffer region width $\delta$,
the $1/\nu$ transport magnitude $\epsilon_{\mathrm{eff}}^{3/2}$ for constructed configurations is found numerically to be proportional
to toroidal flux, as expected from the analytic calculation in section \ref{sec:eps_eff}.
}
\label{fig:s_scaling_of_eps_eff}
\end{figure}

This predicted scaling is borne out numerically, as shown in figure \ref{fig:s_scaling_of_eps_eff}.
There, the numerical construction is carried out for several values of the buffer region size $\delta$ and the boundary aspect ratio $A$. For each case, the equilibrium is
computed using the VMEC code, and then the radial profile of $\epsilon_{\mathrm{eff}}^{3/2}$ is evaluated using the NEO code \citep{nemov-1999}.
The numerical results show a scaling
$\epsilon_{\mathrm{eff}}^{3/2} \propto \psi \propto \epsilon^2$ as expected.

\begin{figure}
\centering
\includegraphics[width=5in]{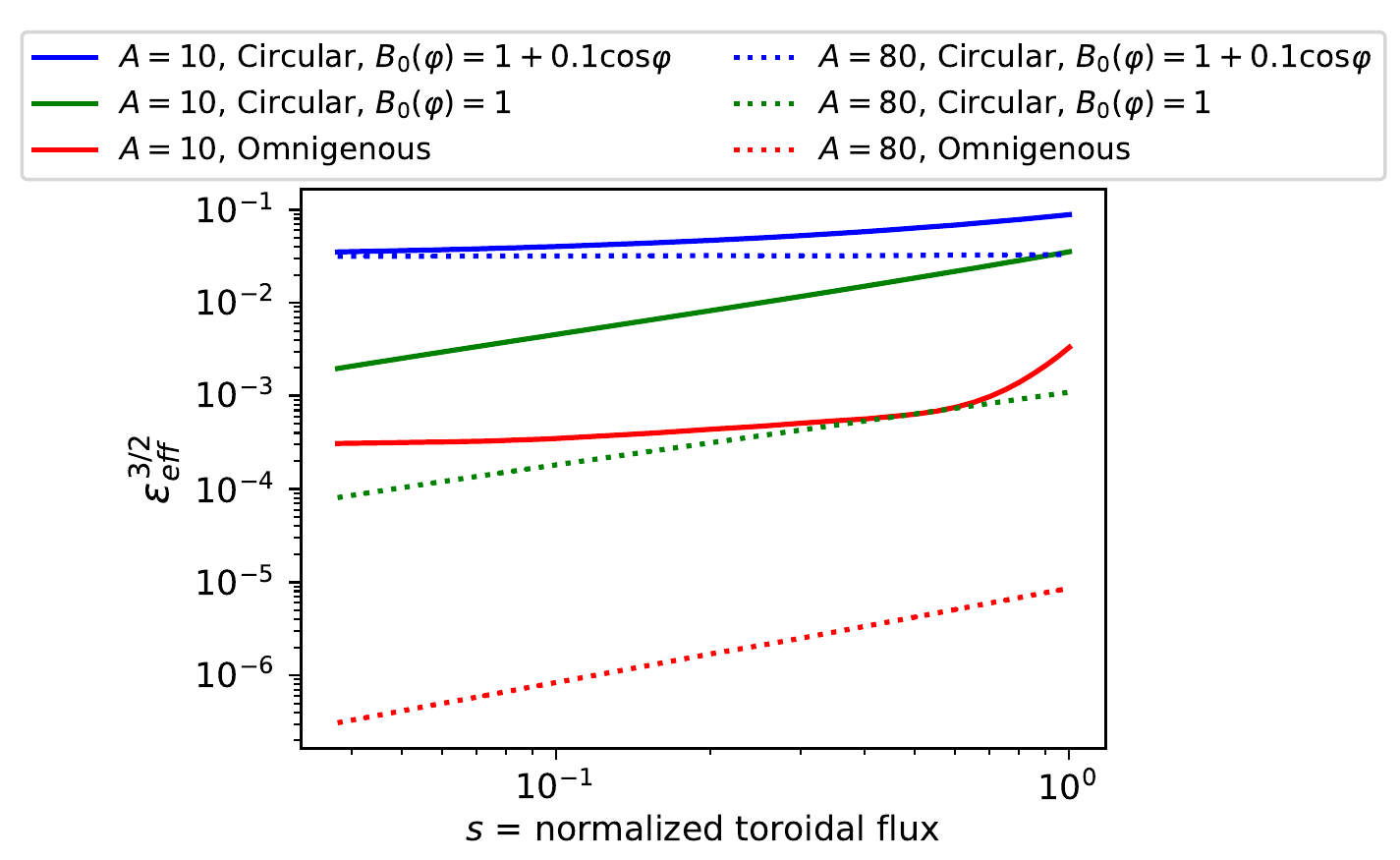}
\caption{For given aspect ratio $A$, the omnigenous construction (red) results in
lower $1/\nu$ transport magnitude $\epsilon_{\mathrm{eff}}^{3/2}$ compared to configurations with the same magnetic axis shape but circular cross-section (green and blue).
}
\label{fig:compareOmniVsCircularXSection}
\end{figure}

Figure \ref{fig:compareOmniVsCircularXSection} demonstrates that the construction for omnigenity results in reduced $\epsilon_{\mathrm{eff}}$ compared to non-optimized configurations of otherwise similar geometry. In particular, we compare the configuration of section \ref{sec:example}
to configurations with the same magnetic axis shape but circular cross-section in the plane perpendicular to the axis. We consider two types of these latter configurations. In the first, shown in green in figure \ref{fig:compareOmniVsCircularXSection}, the radius of the circular cross-sections is independent of toroidal angle, leading to (nearly) constant $B$ along the axis. In the second type of non-optimized configuration, shown in blue in figure \ref{fig:compareOmniVsCircularXSection}, the radius of the circular cross-sections is made to vary with toroidal angle as $\propto 1/\sqrt{1+0.1 \cos\varphi}$, so $B_0(\varphi)$ is matched to that of the constructed omnigenous configurations. Results are shown for two values of aspect ratio $A$, 10 (solid curves) and 80 (dashed). For each configuration, a numerical equilibrium is computed with the VMEC code and $\epsilon_{\mathrm{eff}}^{3/2}$ is then computed using the NEO code \citep{nemov-1999}. At each value of aspect ratio, the constructed omnigenous configuration has smaller $\epsilon_{\mathrm{eff}}^{3/2}$ than either of the non-optimized configurations. Thus, the procedure of section \ref{sec:numerical-solutions} does appear to be a practical way to generate finite-aspect-ratio configurations with reduced $1/\nu$ transport.

\section{Conclusion}\label{sec:conclusion}

We have demonstrated that it is possible to directly construct approximately quasi-isodynamic magnetic equilibria near the magnetic axis, with low computational cost, as compared to conventional optimization.  These solutions are valid to first order in the distance from the magnetic axis, and also satisfy omnigenity (zero bounce-averaged radial drift) at that order for all particle orbits except a small fraction that have bounce points in the neighborhood of the point of maximum magnetic field strength.  This unconfined fraction, $f_{\mathrm{no}}$, can, in principle, be made arbitrarily small, at the cost of introducing sharp behavior in the magnetic field.  It is however unnecessary to make $f_{\mathrm{no}}$ much smaller than the square root of the plasma collisionality, since the effective scattering frequency into and out of the unconfined region in velocity space is proportional to $f_{\mathrm{no}}^2$. 

Our findings imply that quasi-isodynamic fields have the only possible magnetic-field-strength topology that can be achieved near the magnetic axis that satisfies omnigenity while breaking quasi-symmetry.  The present work therefore naturally complements the quasi-axial and quasi-helical cases explored in the two previous papers of this series \citep{landreman-sengupta, landreman-sengupta-plunk}, giving a comprehensive set of tools for constructing stellarator equilibria optimized for collisionless particle confinement near the magnetic axis -- recall that quasi-poloidal symmetry is not possible.  A noteworthy advantage of quasi-isodynamic construction is the additional freedom, affording a much bigger space of possible configurations than can be achieved with quasi-symmetric construction.  With the latter case, the magnetic field strength must be constant on the magnetic axis, and the main freedom is in the shape of the magnetic axis.  With a quasi-isodynamic construction, the on-axis field strength is a free function, and there are two additional free functions of the toroidal angle, corresponding (roughly speaking) to the angle of poloidal rotation of the elliptical cross section ($\alpha(\varphi)$) and elongation of the ellipse ($d(\varphi)$), which must only satisfy certain symmetry requirements.  

We have found that the case of constant magnetic field strength is theoretically forbidden for quasi-isodynamic fields.  Correspondingly, in our numerical constructions, we encountered a limit to how small the magnetic mirror amplitude could be made.  This demonstrates a certain separation between optimization lines, \ie local optimization would not likely stumble upon a quasi-axisymmetric configuration, while searching in the neighborhood of a quasi-isodynamic one.

We studied the $1/\nu$ neoclassical transport associated with some examples of equilibria obtained numerically, and confirmed the expected theoretical scaling of that transport in the distance from the magnetic axis.  We noted that quasi-symmetry here seems to have an advantage due to the trapped particle fraction tending to zero on the magnetic axis -- the magnetic field cannot vary in $\varphi$ there -- so that $1/\nu$ transport (determined solely by trapped particles) tends more sharply to zero toward the magnetic axis.  Both quasi-symmetric or quasi-isodynamic solutions, however, can be constructed with sufficiently small transport to give good (fusion relevant) stellarator design candidates.

To satisfy a certain periodicity condition in our solutions, it was necessary to introduce locations where the geometric condition of omnigenity is broken.  This was done at locations of maximum magnetic field strength, to ensure that all but the marginally trapped particles remain well-confined.  We note however that the perturbation in the magnetic field strength goes to zero at these locations, mitigating the effect of the non-omnigenous component of the magnetic field on marginally trapped particles.  Note that, generally, there exist very weakly trapped particles that drift a macroscopic distance during a bounce time, spending disproportionate time in the neighborhood of the maximum magnetic field, but such particles cannot benefit from the cancellation of radial drift at the two ends of the magnetic well, afforded by the condition omnigenity as applied here.  Perfectly confining them therefore requires a different strategy, namely satisfying zero radial drift locally, \ie at points close to the location of maximum B (note that ${\bf v}_d\cdot\bnabla\psi = 0$ exactly at the maximum).  However, as discussed previously, it is unnecessary to perfectly confine these particles.

We note that it appears possible to achieve smallness of the regions of unconfined orbits without incurring a cost in the smoothness of the magnetic field, if the rotational transform is close to an integer.  It is a curious coincidence that Wendelstein 7-X stellarator does indeed have a near-unity rotational transform, though it was motivated by a different design principle (the island divertor concept).

Possible future continuation of work includes a stellarator design study, where traditional optimization methods are applied with initial states given by our directly constructed solutions; such an approach has the advantage of drastically reducing the parameter space that needs to be searched.  It may also be possible to extend the expansion to higher order in the distance from the magnetic axis, though this would also require extension of the expansion of the CS map, as performed in section \ref{sec:omnigenity-near-qs}.

Acknowledgments.  This work was supported by a grant from the Simons Foundation (560651, ML, PH).

\bibliographystyle{unsrtnat}
\bibliography{omnigenity-near-axis}

\end{document}